\documentclass[fleqn,usenatbib]{mnras}

\usepackage{newtxtext,newtxmath}
\usepackage[T1]{fontenc}

\DeclareRobustCommand{\VAN}[3]{#2}
\let\VANthebibliography\thebibliography
\def\thebibliography{\DeclareRobustCommand{\VAN}[3]{##3}\VANthebibliography}

\usepackage{graphicx}	% Including figure files
\usepackage{amsmath}	% Advanced maths commands
\usepackage{hyperref}
\usepackage{cprotect}
\usepackage{multirow}

\title[Hydrogen-silicate miscibility in sub-Neptunes]{Redefining interiors and envelopes: hydrogen-silicate miscibility and its consequences for the structure and evolution of sub-Neptunes}

\author[Rogers,  Young \& Schlichting]{
James G. Rogers$^{1}$\thanks{E-mail: jr2011@cam.ac.uk}, Edward D. Young$^{2}$ \& Hilke E. Schlichting$^{2}$
\\
$^{1}$Institute of Astronomy, University of Cambridge, Madingley Road, Cambridge CB3 0HA, United Kingdom\\
$^{2}$Department of Earth, Planetary, and Space Sciences, The University of California, Los Angeles, 595 Charles E. Young Drive East, Los Angeles, CA 90095, USA\\
}

\date{Accepted XXX. Received YYY; in original form ZZZ}

\pubyear{\the\year{}}

\begin{document}
\label{firstpage}
\pagerange{\pageref{firstpage}--\pageref{lastpage}}
\maketitle

\begin{abstract}
We present the first evolving interior structure model for sub-Neptunes that accounts for the miscibility between silicate magma and hydrogen. Silicate and hydrogen are miscible above $\sim 4000$~K at pressures relevant to sub-Neptune interiors. Using the H$_2$-MgSiO$_3$ phase diagram, we self-consistently couple physics and chemistry to determine the radial extent of the fully miscible interior. Above this region lies the envelope, where hydrogen and silicates are immiscible and exist in both gaseous and melt phases. The binodal surface, representing a phase transition, provides a physically/chemically informed boundary between a planet’s ``interior'' and ``envelope''. We find that young sub-Neptunes can store several tens of per cent of their hydrogen mass within their interiors. As the planet cools, its radius and the binodal surface contract, and the temperature at the binodal drops from $\sim 4000$ to $\sim 3000$~K. Since the planet's interior stores hydrogen, its density is lower than that of pure-silicate. Gravitational contraction and thermal evolution lead to hydrogen exsolving from the interior into the envelope. This process slows planetary contraction compared to models without miscibility, potentially producing observable signatures in young sub-Neptune populations. At early times ($\sim 10$–$100$~Myr), the high temperature at the binodal surface results in more silicate vapour in the envelope, increasing its mean molecular weight and enabling convection inhibition. After $\sim$~Gyr of evolution, most hydrogen has exsolved, and the radii of miscible and immiscible models converge. However, the internal distribution of hydrogen and silicates remains distinct, with some hydrogen retained in the interior.
\end{abstract}


%%%%%%%%%%%%%%%%%%%%%%%%%%%%%%%%%%%%%%%%%%%%%%%%%%

%%%%%%%%%%%%%%%%% BODY OF PAPER %%%%%%%%%%%%%%%%%%

\section{Introduction} \label{sec:intro}
Sub-Neptunes have sizes between $\sim 1.8-4$~R$_\oplus$, masses between $\sim 2-20$~M$_\oplus$, and orbital periods between $\sim 1-100$~days \citep[e.g.][]{Howard2012,Fressin2013,Petigura2013,Wu_Lithwick2013,Wolfgang2016,Chen2017}. Sub-Neptunes are separated from their smaller counter-parts, the super-Earths, by the ``radius valley'', a paucity in planet occurrence extending through parameter space \citep[e.g.][]{Fulton2017,VanEylen2018,Petigura2022}. It is commonly thought that many super-Earths and sub-Neptunes were formed as a single progenitor population of silicate-rich interiors with H/He-rich envelopes. Then, stellar-driven atmospheric escape stripped the envelopes from planets with lower masses and smaller orbital separations to form the population of super-Earths \citep[e.g.][]{Owen2013,LopezFortney2013,Ginzburg2018,Gupta2019}. Evidence in favour of this scenario comes from multiple observational channels, including direct observations of atmospheric escape from sub-Neptunes \citep[e.g.][]{DosSantos2023a,Loyd2025}, agreement between atmospheric escape theory and observations regarding radius valley trends as a function of orbital period and stellar mass \citep[e.g.][]{Owen2017,Gupta2020}, and populations of inflated young sub-Neptunes \citep[e.g.][]{Vach2024a,Fernandes2025,Rogers2025b}.

% The existence of a model for sub-Neptune structure and evolution that can successfully reproduce observations implies that it can be used to infer observed planet properties that are traditionally unobservable. Examples include inferences of interior composition, initial and present envelope mass fractions, and ``core'' mass \citep[e.g.][]{OwenMorton2016,Wu2019,Rogers2021,Rogers2023}. From these inference studies, it is believed that most sub-Neptunes host a H/He envelope mass fraction of order $\sim 1$\%, and have interior compositions consistent with Earth. However, the underlying models used in these inferences are typically simple physical models that ignore important chemical effects.

Unlike gas giants, whose mass budgets are dominated by their large H/He-dominated envelopes, sub-Neptunes store most of their mass in their rocky interiors. As a result, the silicate reservoir dictates much of the chemistry acting inside a sub-Neptune. This can cause composition and structural changes as a result of envelope gases reacting with underlying magma-oceans \citep[e.g.][]{Chachan2018,Kite2020,Kite2021,Schlichting2022,Charnoz2023,Rogers2024b,Shorttle2024,Werlen2025,Werlen2025b}. Studies have also shown that silicate vapour, originating from a magma ocean, can alter the envelope structure via the introduction of mean molecular weight gradients. Whereas simple sub-Neptune models without chemistry predict a convective hydrogen-rich envelope in contact with a magma ocean, mean molecular weight gradients can inhibit convection, and ultimately change the predicted radius of a sub-Neptune for a given envelope mass fraction \citep[e.g.][]{Leconte2017,Brouwers2020,Ormel2021,Misener2022,Vazan2023,Steinmeyer2024,Vazan2024}. 
% The implication here is that we may be underestimating the hydrogen content of sub-Neptunes with the use of simple models.

A phenomenon that remains under-explored in the context of sub-Neptunes is that of miscibility. To date, all sub-Neptune evolution models assume a magma ocean surface, which separates a silicate-rich interior from a H/He-rich envelope. The pressure and temperature at this interior-envelope boundary dictates the chemistry at play. Gases are typically allowed to dissolve inside the magma ocean assuming a solubility law, and silicate vapour can evaporate into the envelope. These build upon experimental studies showing that hydrogen, among other gases, can dissolve into a magma ocean at high temperatures and pressures \citep[e.g.][]{Shinozaki2014,Horn2023a}. However, \textit{ab initio} calculations have shown that silicate melt and hydrogen are miscible above temperatures of $\sim 4000$~K \citep{Young2024,Stixrude2025,Young2025}. Miscibility implies that both chemical components merge to form a single, supercritical, homogenous mixture, that can mix in any proportion \citep[see discussion in][]{Markham2022}. Since simple, chemistry-free models suggest that temperatures at interior-envelope boundaries can exceed $\sim10,000$~K at early evolutionary phases, miscibility is therefore expected to significantly alter a sub-Neptune's structure and evolution. Crucially, even models that consider solubility at a magma ocean surface with temperatures $\lesssim 4000$~K will have missed the consequences of a phase change occurring at the boundary between miscible and immiscible regions deeper within the planet.

In this study, we introduce an evolving interior structure model for sub-Neptunes that accounts for hydrogen-silicate miscibility. Our paper is laid out as follows: in Section \ref{sec:method} we summarise the relevant chemistry and physics required to model sub-Neptunes with miscible interiors. We introduce our numerical model and present results in Section \ref{sec:results}, followed by discussion and conclusions in Sections \ref{sec:discussion} and \ref{sec:conclusion}, respectively.

\section{Method} \label{sec:method}
We seek to construct spherically-symmetric 1-D models of sub-Neptunes in hydrostatic and thermochemical equilibrium, accounting for the miscibility between hydrogen and silicate melt. Here we provide a brief primer on phase equilibria, and show how we incorporate this framework into our planetary structure models.

\subsection{Chemical phase equilibria} \label{sec:phase_equilibria}
Miscibility between two chemical components occurs when their intermolecular forces are similar in magnitude and length scale, allowing the two components to form a single homogenous mixture. This is in contrast to a gas dissolving into a fluid, in which case the fluid can become saturated. Supercritical miscible mixtures, on the other hand, can mix in any proportion. Our goal here is to utilise the phase diagram for the hydrogen-silicate system and find the phase boundary which demarcates regions of parameter space (namely, temperature, pressure and mole fractions) for which silicate melt and hydrogen become miscible. We begin with the enthalpy, $H$, of a system:
\begin{equation}
    H = U - PV,
\end{equation}
where $U$ is the internal energy, $P$ is the pressure and $V$ is the volume. At constant pressure, $\Delta H$ is the energy released or absorbed during a chemical reaction. From this, we define the Gibbs free energy:
\begin{equation}
    G = H - TS,
\end{equation}
where $S$ is the entropy. The Gibbs free energy is the energy available for non-mechanical work of a system, such as chemical reactions. At fixed $P$ and $T$, chemical reactions proceed in the direction of lower Gibbs free energy. From this one can define the chemical potential as the derivative of the Gibbs free energy with respect to composition. To avoid confusion with mean molecular weight, $\mu$, in this paper we denote the chemical potential as $\psi_i$ for chemical component $i$:
\begin{equation} \label{eq:chemical_potential}
    \psi_i = \bigg( \frac{\partial \hat{G}}{\partial x_i} \bigg)_{P, T, x_{j, j \neq i}},
\end{equation}
where $\hat{G}$ is the Gibbs free energy per mole, and $x_i$ is the mole fraction of component $i$. \textit{In thermochemical equilibrium, chemical potentials for every component across all phases are equal.} Thus, in order to specify the conditions for chemical equilibrium, we calculate partial derivatives of Gibbs free energies and find mole fractions for which the derivatives with respect to mole fractions at a specified $T$ and $P$ are equal. 

\subsubsection{Chemical mixtures}
The change in Gibbs free energy due to mixing chemical components is:
\begin{equation}
    \Delta \hat{G}_\text{mix} = \Delta \hat{H}_\text{mix} - T \Delta \hat{S}_\text{mix},
\end{equation}
where all extensive properties are now molar, denoted with hats, and ``mix'' subscripts refer to the changes  due to mixing. An \textit{ideal mixture} is defined as one in which $\Delta \hat{H}_\text{mix}=0$. In this case, the change in Gibbs free energy due to mixing becomes:
\begin{equation} \label{eq:deltaGmix_generic}
    \Delta \hat{G}_\text{ideal mix} = RT \sum_i x_i \ln x_i,
\end{equation}
where $R$ is the gas constant and we have used $\Delta \hat{S}_\text{mix} = -R \sum_i x_i \ln x_i$, derived from statistical mechanics due to the available configurations of the system.

For our system, we have two components; hydrogen and silicate, and two available phases; melt and gas (or the fluid equivalent). We denote mole fractions for these components as $x_{\text{H}_2}$ and $x_\text{sil}$, respectively. In a two component system, referred to as a binary mixture, we have that $x_{\text{H}_2} = 1 - x_\text{sil}$. When denoting components in a particular phase, we adopt the following notation, e.g. $x_{\text{H}_2}^\text{melt}$ and $x_{\text{H}_2}^\text{gas}$ for the mole fractions of hydrogen in the melt and gas phases, respectively. From Equation \ref{eq:deltaGmix_generic}, the change in Gibbs free energy for our binary ideal mixture is:
\begin{equation} \label{eq:DeltaG_ideal}
    \Delta \hat{G}_\text{ideal mix} = RT \bigg[ (1 - x_{\text{H}_2}) \ln (1 - x_{\text{H}_2}) + x_{\text{H}_2} \ln x_{\text{H}_2} \bigg].
\end{equation}

\begin{figure}
	\includegraphics[width=\columnwidth]{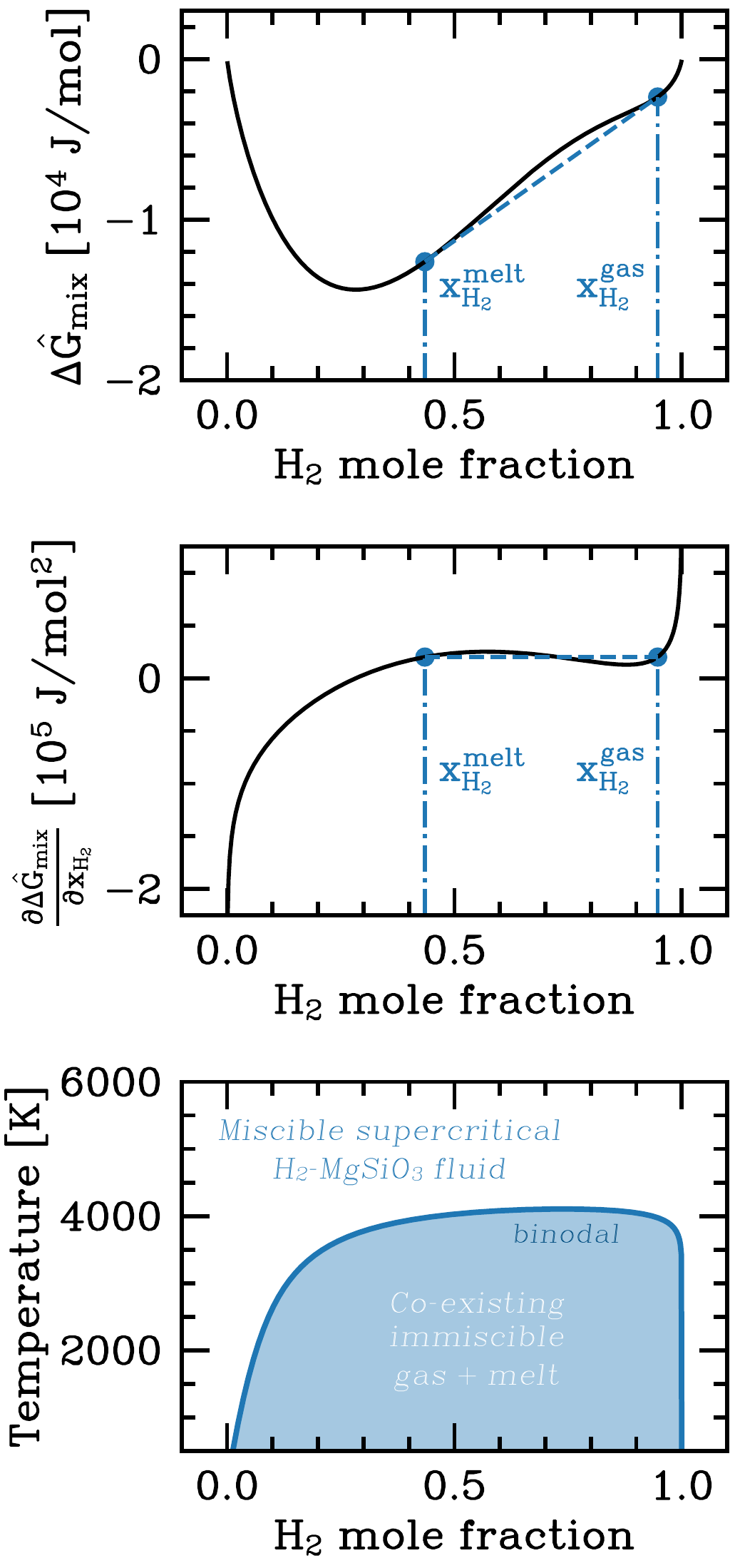}
    \centering
        \cprotect\caption{Upper panel: The change in Gibbs free energy of mixing, $\Delta \hat{G}_\text{mix}$, at $3975$~K and $1$~GPa for a binary, non-ideal mixture of hydrogen and MgSiO$_3$ (Equation \ref{eq:DeltaG_mix}) as a function of H$_2$ mole fraction. Conditions for co-existence of hydrogen in the gas and melt phases exist where the gradient of $\Delta \hat{G}_\text{mix}$ (equivalent to the chemical potential, as shown in the second panel) are equal and positive, as highlighted by the blue dashed line. The hydrogen mole fractions in the melt, $x_{\text{H}_2}^\text{melt}$, and gas phases, $x_{\text{H}_2}^\text{gas}$, are highlighted by blue dot-dashed lines. Lower panel: the binodal, also known as a co-existence curve, demarcates the boundary between miscible and immiscible regions of parameter space. This is shown at $1$~GPa. For temperatures above the binodal, hydrogen and MgSiO$_3$ melt form a homogenous miscible fluid. For temperatures below the binodal, hydrogen and MgSiO$_3$ exist in two phases: gas, including H$_2$, SiO, Mg and O$_2$, and silicate melt (rain) with hydrogen dissolved inside.} \label{fig:DeltaG} 
\end{figure} 

In this study, we account for non-ideal mixing effects with non-zero changes in enthalpy with mixing (e.g. $\Delta \hat{H}_\text{mix} \neq 0$). We make use of the results of \citet{Stixrude2025}, in which Density Functional Theory (DFT)-molecular dynamics simulations of the H$_2$-MgSiO$_3$ binary system were performed. They introduced an asymmetric, non-ideal term to Equation \ref{eq:DeltaG_ideal}:
\begin{equation} \label{eq:DeltaG_mix}
\begin{split}
\Delta \hat{G}_\text{mix}  & = RT \bigg[ (1 - x_{\text{H}_2}) \ln (1 - x_{\text{H}_2}) + x_{\text{H}_2} \ln x_{\text{H}_2} \bigg] \\
 & +  (1 - x_{\text{H}_2})x_{\text{H}_2}(A(1 - x_{\text{H}_2}) + Bx_{\text{H}_2}) \bigg( 1 - \frac{T}{C} + \frac{P}{D} \bigg).
\end{split}
\end{equation}
Here, $P$ is the pressure in GPa, and $A=-6.26 \times 10^3$, $B=7.86\times 10^{5}$, $C = 4.67 \times 10^3$ and $D = -35.0$ are parameters fit to the DFT simulations \citep{Stixrude2025}.

The upper panel of Figure \ref{fig:DeltaG} shows the change in molar Gibbs free energy from non-ideal mixing $\Delta \hat{G}_\text{mix}$ from Equation \ref{eq:DeltaG_mix} at $3975$~K and $1$~GPa. As previously stated, chemical potentials, or composition derivatives of Gibbs free energies, for every phase of a component are equal in thermochemical equilibrium. Thus, for a given temperature and pressure, one determines the mole fractions of H$_2$ in the melt and gas phase, $x_{\text{H}_2}^\text{melt}$ and $x_{\text{H}_2}^\text{gas}$, respectively, by locating the values of $x_{\text{H}_2}$ for which the derivative of $\Delta \hat{G}_\text{mix}$ are equal. These are highlighted in the top panel of Figure \ref{fig:DeltaG} as blue circles joined by their mutual tangent. The chemical potential is shown in the middle panel. We also require that the second derivative of the Gibbs free energy at these locations to be positive for a stable mixture.\footnote{Metastable mixtures can also exist where the second derivative of Gibbs free energy of mixing is not positive. In this case, phase separation can occur spontaneously without the need for nucleation. This is often referred to as \textit{spinodal} decomposition.}

\subsubsection{The binodal surface}
The \textit{binodal}, also referred to as a co-existence curve,\footnote{Occasionally, the binodal is also referred to as a ``solvus'', however, this is typically in reference to a solid-solid mixture, such as in alloys or mineral exsolution.} demarcates regions of parameter space where a binary mixture is miscible or immiscible. For the H$_2$-MgSiO$_3$ system, it is calculated by determining the mole fractions of hydrogen and silicate coexisting in the melt and gas phases as a function of temperature and pressure (as previously described). An example is shown in the bottom panel of Figure \ref{fig:DeltaG} at a constant pressure of $1$~GPa. For temperatures below the binodal, hydrogen and silicate are immiscible and co-exist in two distinct phases; gas and melt. In a practical sense, this should be interpreted as a mixture of hydrogen and silicate in the gas phase (with the silicate forming ``silicate vapour'' species such as SiO), as well as liquid droplets of silicate melt with hydrogen dissolved inside. For temperatures above the binodal, hydrogen and silicate melt are completely miscible and form a homogenous mixture.

An important variable within our planetary model is the mass fraction of H$_2$ in phase $\alpha$ contained inside each spherical mass shell, defined as $X_{\text{H}_2}^\alpha \equiv M_{\text{H}_2}^\alpha / (M_{\text{H}_2}^\alpha + M_\text{sil}^\alpha)$, where $M_{\text{H}_2}^\alpha$ and $M_\text{sil}^\alpha$ are the masses of hydrogen and silicate (in phase $\alpha$) within the shell. From the binodal, we can calculate this quantity, specifically for the gas phase:
\begin{equation}
    X_{\text{H}_2}^\text{gas} = \frac{x_{\text{H}_2}^\text{gas} \, \mu_{\text{H}_2}}{x_{\text{H}_2}^\text{gas} \, \mu_{\text{H}_2} + x_{\text{sil}}^\text{gas} \, \mu_{\text{sil}}},
\end{equation}
where $\mu_{\text{H}_2} = 2.016$~g mol$^{-1}$ and $\mu_{\text{sil}} = 100.39$~g mol$^{-1}$ are the mean molecular weights of H$_2$ and MgSiO$_3$, respectively. For a given pressure and temperature, the above framework allows us to determine the chemical composition of a sub-Neptune throughout its interior and envelope. 

The lower panel of Figure \ref{fig:DeltaG} demonstrates the sensitivity of the binodal surface as a function of hydrogen composition. For hydrogen mole fractions close to zero or unity, the temperature drops rapidly for a small change in composition. At intermediate hydrogen mole fractions, a small change in temperature dramatically changes the composition. These properties make solving for mole fractions from the Gibbs free energies computationally non-trivial. To alleviate these complexities, we use analytic functional fits to the binodal surface, as described in Appendix \ref{app:binodal}.

\subsection{Planetary structure model} \label{sec:planetary_structure}
\begin{figure}
	\includegraphics[width=\columnwidth]{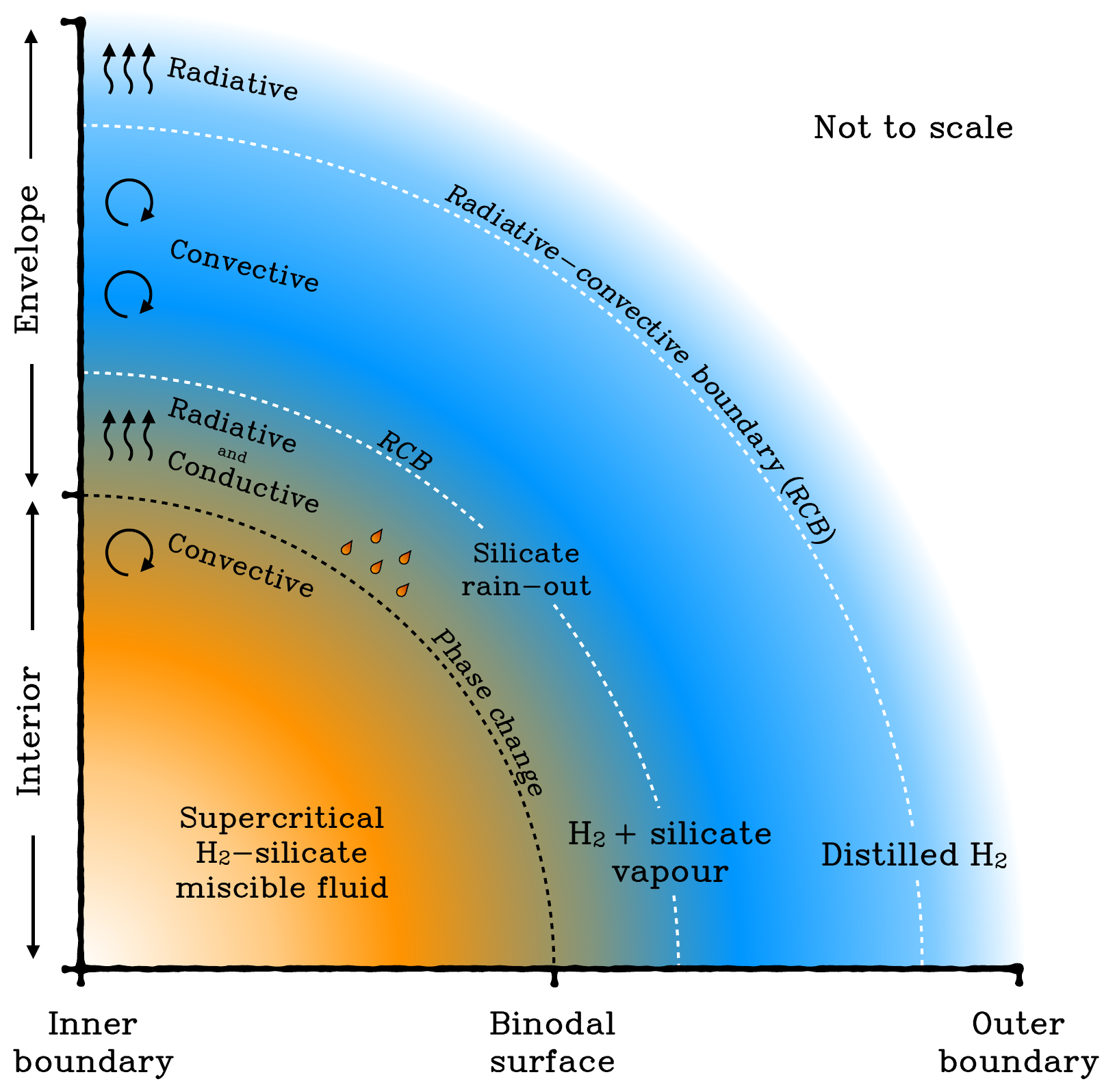}
    \centering
        \cprotect\caption{A schematic for the interior structure of sub-Neptunes. Moving radially outwards from the planetary centre: the ``interior'' is defined as the region interior to the binodal surface. At the binodal, a phase change occurs as the convective, miscible, hydrogen-silicate fluid speciates into gas and melt phases. The region above the binodal is defined as the ``envelope''. The silicate vapour introduces a mean molecular weight gradient, which can inhibit convection. In this case, heat is transported via conduction and radiative diffusion. Silicate-rich melt droplets rain-out to rejoin the interior. Continuing to move radially outward, the envelope becomes unstable to convection and the gas become progressively more hydrogen-rich. The very upper region of the envelope is almost pure hydrogen gas, and heat is transported by radiative diffusion. Radiative-convective boundaries (RCBs) are shown as white dashed lines.} \label{fig:Schematic} 
\end{figure} 
In our sub-Neptune models, we define the ``interior'' as the region with temperatures and pressures above the binodal surface, implying a miscible hydrogen-silicate fluid. The ``envelope'' is then defined as the outer region with temperatures and pressures below the binodal surface. In the envelope, hydrogen and silicate co-exist in the gas and melt (rain) phases. The general structure is presented as a schematic in Figure \ref{fig:Schematic} and moves beyond the notion that sub-Neptunes have fixed surfaces, often referred to as ``magma ocean surfaces''. Instead, the binodal surface provides a convenient demarcation between material properties within the planet due to a chemical phase change. The radial position of the binodal will not be constant over a planet's lifetime as the phase change front evolves with time.

As argued in \citet{Schlichting2022,Young2024,Rogers2024b,Young2025}, the presence of hydrogen in a silicate-rich interior may affect the buoyancy of iron droplets required for the formation of a differentiated iron core via iron rain-out. For this reason, and for the sake of simplicity in this initial study, we choose not to include iron in our planet models. Here, we only consider the binary mixture of H$_2$-MgSiO$_3$, and leave the addition of further chemical components for future work. We also ignore the presence of helium, since the phase equilibria of the H$_2$-He-MgSiO$_3$ system has not been explored to date.

To build a sub-Neptune model, we must solve the following differential equations. Firstly, for mass conservation:
\begin{equation} \label{eq:mass_cons}
    \frac{\partial r}{\partial m} = \frac{1}{4 \pi r^2 \rho},
\end{equation}
where $\rho$ is the density and $m$ is the mass contained within radius $r$. Then, for hydrostatic equilibrium:
\begin{equation} \label{eq:hydro_eq}
    \frac{\partial P}{\partial m} = -\frac{Gm}{4 \pi r^4},
\end{equation}
where $G$ is the gravitational constant. Finally, for heat transport:
\begin{equation} \label{eq:heat_transport}
    \frac{\partial T}{\partial m} = -\frac{Gm}{4 \pi r^4} \frac{T}{P} \nabla,
\end{equation} 
where $\nabla \equiv \partial \ln T / \partial \ln P$ is the temperature gradient and depends on the material properties.
% \begin{equation} \label{eq:fH2_distil}
%     \frac{\partial f_{H_2}^\text{gas}}{\partial m} = \bigg ( \frac{\partial f_{H_2}^\text{gas}}{\partial T} \bigg)_P \, \frac{\partial T}{\partial m} + \bigg ( \frac{\partial f_{H_2}^\text{gas}}{\partial P} \bigg)_T \, \frac{\partial P}{\partial m},
% \end{equation}
% which is only solved in the envelope where the chemical components decompose into gas and melt phases.
For each planetary model, we choose a planet mass, $M_\text{p}$, a global hydrogen mass fraction, $X_{H_2} = M_{H_2} / M_\text{p}$, where $M_{H_2}$ is the total mass of hydrogen within the planet, and an equilibrium temperature, $T_\text{eq}$. As adopted in many studies, we assume the planet is in quasi-equilibrium at each evolutionary snapshot (e.g. $\partial / \partial t$ terms are ignored), which is valid if evolutionary timescales are sufficiently long when compared to the planet's age \citep[e.g.][]{Hubbard1977,Piso2014,Lee2014}. We then connect these snapshots together according to the conservation of energy in order to model evolution, as discussed in Section \ref{sec:evolution}. A corollary of this assumption is that the internal luminosity is constant throughout the planet. We set this luminosity as an additional free parameter, denoted as the luminosity at the upper-most radiative-convective boundary, $L_\text{rcb}$. We confirm \textit{a posteriori} that an assumption of constant luminosity is valid in Section \ref{sec:convectionInhib}.

\subsubsection{Interior material properties} \label{sec:int_prop}
For regions of the planet with temperatures and pressures above the binodal surface, hydrogen and silicate melt (MgSiO$_3$ in our case) are completely miscible and form a homogenous mixture. The presence of hydrogen in the silicate melt reduces its density, creating a puffier interior than in standard models. We assume the interior is fully convective, meaning that the miscible silicate and hydrogen are perfectly mixed. As a result, we assume a constant hydrogen mass fraction throughout the entire interior, denoted as $X_{\text{H}_2\text{,int}}$, which we solve for as part of our numerical procedure. To calculate the interior density, we assume an ideal mixture of hydrogen and silicate melt with use of a harmonic mean molar volume, which has been found to best-fit results from DFT simulations \citep{Young2025}. The density of the uncompressed H$_2$-MgSiO$_3$ mixture is given by:
\begin{equation} \label{eq:Vmix}
    \rho_{\text{mix},0} = (x_{\text{H}_2} \mu_{\text{H}_2} + x_{\text{sil}} \mu_{\text{sil}}) \, \bigg(x_{\text{H}_2} \frac{\rho_{\text{H}_2,0}}{\mu_{\text{H}_2}} + x_{\text{sil}} \frac{\rho_{\text{sil},0}}{\mu_{\text{sil}}} \bigg)
\end{equation}
where $\rho_{\text{H}_2,0}$ and $\rho_{\text{sil},0}$ are the uncompressed densities of pure hydrogen and silicate. For the compression of the MgSiO$_3$ melt, we use the Vinet fit to the \citet{deKoker2009} equation of state from \citet{Wolf2018}. For hydrogen, we use the equation of state from \citet{Chabrier2019}. For both species, we pre-calculate a grid of material properties and interpolate during the numerical procedure (Section \ref{sec:numerical_proc}). 

We assume the interior is fully convective with an adiabatic temperature gradient:
\begin{equation}
    \nabla_\text{ad} = \bigg ( \frac{\partial \ln T}{\partial \ln P}\bigg)_S = \frac{P \delta}{T \rho c_P},
\end{equation}
where we assume a constant specific isobaric heat capacity for MgSiO$_3$ melt of $c_P = 1.195 \times 10^3$~J K$^{-1}$ kg$^{-1}$, and $\delta$ is a measure of thermal expansivity:
\begin{equation}
    \delta \equiv - \bigg ( \frac{\partial \ln \rho}{\partial \ln T}\bigg)_P,
\end{equation}
which is calculated from the \citet{Wolf2018} equation of state tables. Note that both equation of state tables for hydrogen and silicate melt, as well as the DFT molecular dynamic simulations of \citet{Stixrude2025}, account for effects such as molecular dissociation and ionisation, hence these processes are taken into account when calculating physical and chemical properties. We assume the presence of hydrogen does not alter adiabatic temperature gradients or thermal expansivities in the interior. We confirm that the temperatures within the interior of our sub-Neptune models are sufficiently greater than the liquidus of silicate melt, such that we need not consider crystallisation in the planet mass range of $3 \leq M_\text{p} / M_\oplus \leq 12$ considered in this study.

\begin{figure*}
	\includegraphics[width=2.0\columnwidth]{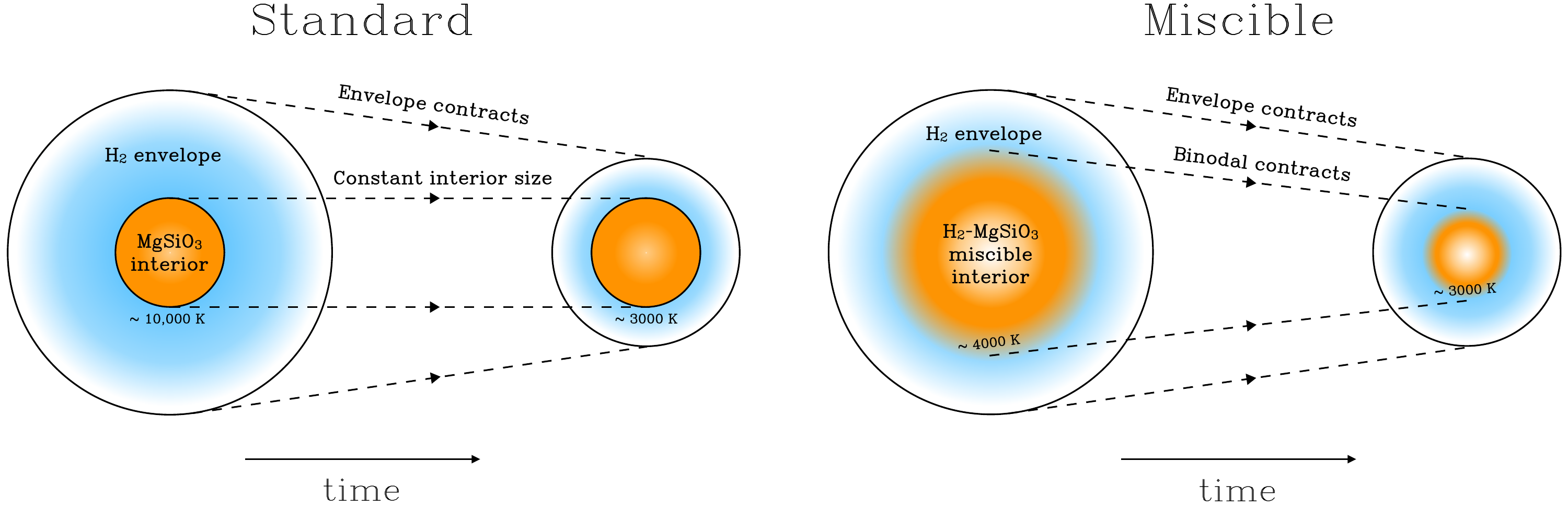}
    \centering
        \cprotect\caption{Schematic showing the evolution of two interpretations of sub-Neptune interiors. In the standard model, a silicate interior is distinct from a hydrogen-rich envelope. As the planet contracts, the interior-envelope boundary does not significantly contract with time. The temperature at this interface can reach of order $\sim 10,000$~K at early ages. In the miscible model, the interior-envelope boundary is defined by a \textit{binodal surface}, which delineates regions in which hydrogen and silicate are miscible or immiscible. The temperature at the binodal surface does not change significantly with time. The radial position of the binodal surface contracts with time.} \label{fig:SchematicEvolution} 
\end{figure*}

\subsubsection{Envelope material properties} \label{sec:env_prop}
The envelope is defined as the region for which temperatures and pressures are below the binodal surface. Hydrogen and silicate decompose into the gas and melt phases. As in the interior, we use the equation of state from \citet{Chabrier2019} for hydrogen. We assume silicate vapour speciates into gaseous Mg, SiO and O$_2$. In the absence of realistic equations of state for these species, we assume the true densities of these high mean molecular components follows the same fractional deviation from the ideal gas law as hydrogen \citep[based on the equation of state from ][]{Chabrier2019}. Mathematically, we state that the gas density for an arbitrary mean molecular weight at a given temperature and pressure is given by:
\begin{equation} \label{eq:rho_nonideal}
    \rho(P,T,\mu) = \frac{\rho_\text{non-ideal}(P,T,\mu_{\text{H}_2})}{\rho_\text{ideal}(P,T,\mu_{\text{H}_2})} \, \rho_\text{ideal}(P,T,\mu),
\end{equation}
where $\rho_\text{ideal}$ is calculated using the ideal gas law and $\rho_\text{non-ideal}$ from the \citet{Chabrier2019} equation of state.

The relative mole fractions of silicate vapour species are calculated with use of the binodal, as described in Section \ref{sec:phase_equilibria}. The mean molecular weight of the gas is calculated as follows:
\begin{equation} \label{eq:mmw_gas}
    \mu_\text{gas} = x_{\text{H}_2}^\text{gas} \, \mu_{\text{H}_2} + \frac{1}{3}x_{\text{sil}}^\text{gas} (\mu_\text{SiO} + \mu_\text{Mg} + \mu_{\text{O}_2}),
\end{equation}
where $\mu_\text{SiO} = 44.08$~g mol$^{-1}$, $\mu_\text{Mg} = 24.31$~g mol$^{-1}$ and $\mu_{\text{O}_2} = 32.00$~g mol$^{-1}$ are the mean molecular weights of Mg, SiO and O$_2$, respectively.

The melt phase corresponds to droplets of silicate melt with dissolved hydrogen gas. We assume perfect rain-out of such droplets, in that they fall under gravity and are heated as they rejoin the miscible interior. Since most of the mass of a droplet is in silicate rather than hydrogen, the rain-out process acts to distil the envelope to become more and more hydrogen rich as temperatures and pressures drop towards the top of the envelope. 

Three heat transport mechanisms can operate in the envelope: convection, conduction and radiative diffusion. Like in the interior, we again assume an adiabatic temperature gradient for convective regions, in this case the moist adiabat from \citet{Graham2021} (their Equation 1) to account for condensible silicate gas species and rain-out. For this, we calculate the latent heat of condensation, $L_\text{cond}$ with enthalpies for MgSiO$_3$ melt, SiO, Mg and O$_2$ from the NIST database. We note that the adiabatic temperature gradient $\nabla_\text{ad}$ from \citet{Graham2021} was derived assuming an ideal gas law. To account for non-ideality, we take a similar approach as in Equation \ref{eq:rho_nonideal} for gas densities. We state that the fractional deviation of the true moist adiabatic temperature gradient from the ideal case of \citet{Graham2021} follows the deviation between ideal and non-ideal hydrogen from \citet{Chabrier2019}. Mathematically:
\begin{equation}
    \nabla_\text{ad} = \frac{\nabla_{\text{ad, H}_2\text{, non-ideal}}}{\nabla_{\text{ad, H}_2\text{, ideal}}} \, \nabla_\text{ad, moist, ideal},
\end{equation}
where $\nabla_{\text{ad, H}_2\text{, non-ideal}}$ comes from \citet{Chabrier2019}, $\nabla_{\text{ad, H}_2\text{, ideal}}=7/5$ and $\nabla_\text{ad, moist, ideal}$ comes from \citet{Graham2021}.

In regions where heat is transported most efficiently by radiative diffusion and conduction, the temperature gradient is given by:
\begin{equation} \label{eq:rad_diff}
    \nabla_\text{rad} = \bigg ( \frac{\partial \ln T}{\partial \ln P}\bigg)_\text{rad} = \frac{3 \kappa_\text{eff} P L_\text{rcb}}{64 \pi G m \sigma T^4},
\end{equation}
where $\sigma$ is the Stefan-Boltzmann constant and $\kappa_\text{eff}$ is an ``effective opacity'' \citep{Vazan2020}:
\begin{equation} \label{eq:kappa_eff}
    \frac{1}{\kappa_\text{eff}} = \frac{1}{\kappa} + \frac{1}{\kappa_\text{c}},
\end{equation}
where $\kappa$ is the Rosseland mean opacity of the gas and $\kappa_\text{c}$ is the conductive opacity, given by:
\begin{equation}
    \kappa_\text{c} = \frac{16 \sigma T^3}{3 \rho \lambda},
\end{equation}
where $\lambda$ is the thermal conductivity. As in \citet{Misener2023}, we take the electrical conductivities for hydrogen from experimental results from \citet{McWilliams2016} and then use the Wiedemann–Franz law to convert these to thermal conductivities. We assert a lower bound of $\lambda = 2 \times 10^5 \text{ erg s}^{-1} \text{ cm}^{-1} \text{ K}^{-1}$, appropriate for nucleic contributions in the parameter space region of interest \citep{French2012}.

\begin{figure*}
	\includegraphics[width=2.0\columnwidth]{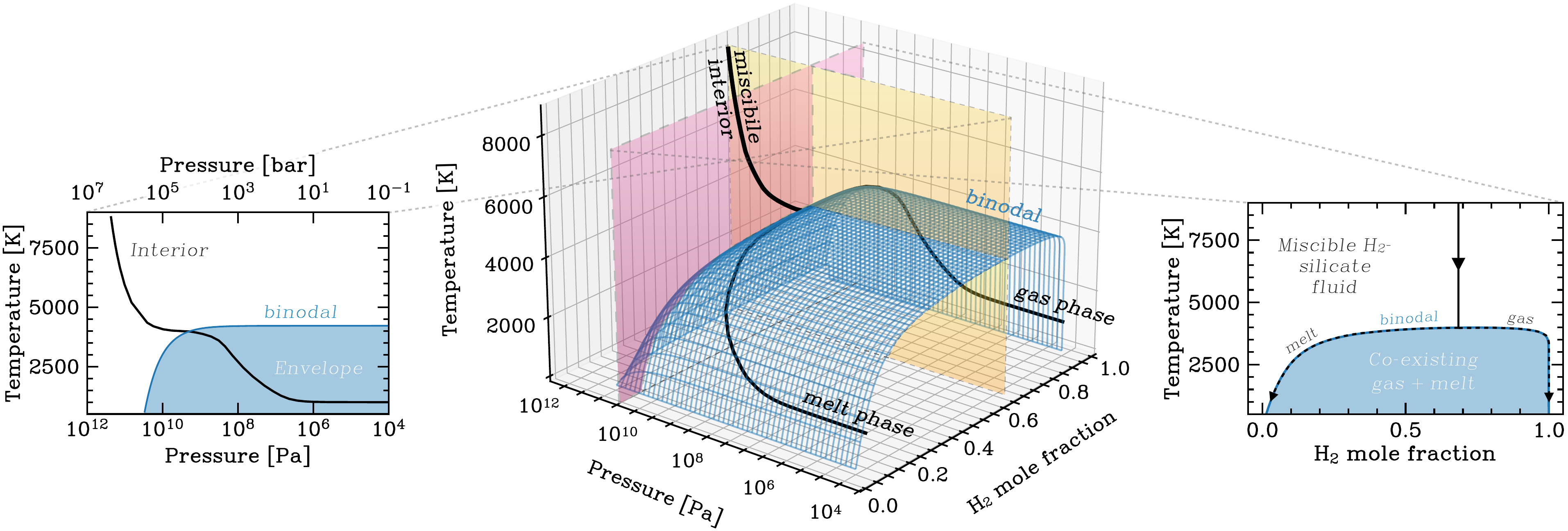}
    \centering
        \cprotect\caption{The interior structure profile for a $6 M_\oplus$ sub-Neptune with equilibrium temperature of $1000$~K and global hydrogen mass fraction of $5$\% is shown as a black line in pressure-temperature-hydrogen mole fraction space. The blue surface represents the binodal, demarcating regions of parameter space where hydrogen and silicate melt are miscible or immiscible. In the interior of the planet, the temperatures and pressures are higher than the binodal, hence, hydrogen and silicate melt are miscible and form a homogenous supercritical mixture. In the envelope, temperatures and pressures are below the binodal, meaning the chemical components speciate into melt and gas phases, with the respective hydrogen mole fractions represented by individual black lines. As pressure and temperature drop, silicate-rich melt rains out, decreasing the hydrogen mole fraction in the melt phase, and increasing the hydrogen mole fraction in the gas phase. Projections of the profile in the pressure-temperature plane and hydrogen mole fraction-temperature plane are shown on the left (yellow projection) and right-hand-side (pink projection), respectively.} \label{fig:3D} 
\end{figure*}

We adopt a power law optical opacity scaling for Solar metallicity gas from \citet{Rogers2010,Owen2017}:
\begin{equation} \label{eq:opacity}
    \kappa = 1.3 \times 10^{-2} \, \bigg( \frac{T}{1000 \, \text{K}} \bigg)^{0.45} \, \bigg( \frac{P}{1 \, \text{bar}} \bigg)^{0.68} \text{ cm}^2 \text{ g}^{-1}.
\end{equation}
We highlight that very little study has been performed on gas opacities in regions of parameter space relevant to the base of sub-Neptune envelopes, where silicate vapour is likely present and will increase the opacity. For now, we simply extrapolate Equation \ref{eq:opacity} for all regions, but note that this topic warrants more study.

To determine whether a region of the envelope is stable to convection, we make use of the Schwarschild criterion, which states that stability occurs when:
\begin{equation}
    \nabla_\text{rad} < \nabla_\text{ad}.
\end{equation}
We also account for convection inhibition due to mean molecular weight gradients following \citet{Markham2021,Markham2022}, in which a critical mole fraction is defined:
\begin{equation}
    x_\text{sil, crit}^\text{gas}= \bigg[ \bigg( \frac{\mu_\text{melt} L_\text{cond}}{RT} - 1\bigg) \bigg( \frac{\mu_\text{gas}}{\mu_{\text{H}_2}} - 1 \bigg) \bigg]^{-1},
\end{equation}
where $\mu_\text{gas}$ comes from Equation \ref{eq:mmw_gas} and $\mu_\text{melt}$ is the mean molecular weight of the condensed silicate-hydrogen melt:
\begin{equation}
    \mu_\text{melt} = x_{\text{H}_2}^\text{melt} \, \mu_{\text{H}_2} + x_{\text{sil}}^\text{melt} \, \mu_{\text{sil}}.
\end{equation}
If $x_\text{sil}^\text{gas} > x_\text{sil, crit}^\text{gas}$, then convection is inhibited and heat is transported by conduction and radiative diffusion.

\subsubsection{Numerical procedure} \label{sec:numerical_proc}
The independant variables for our planetary structure model are: planetary mass, $M_\text{p}$, global hydrogen mass fraction, $X_{H_2}$, equilibrium temperature, $T_\text{eq}$, and the luminosity at the outermost radiative-convective boundary, $L_\text{rcb}$. To solve the equations for planetary structure (Equations \ref{eq:mass_cons}-\ref{eq:heat_transport}), we use a shooting method with an adaptive step \verb|RK45| integrator with relative and absolute tolerances of $10^{-6}$ and $10^{-8}$, respectively. The model is initialised with a guess for the central pressure of the planet, $P_\text{c}$, central temperature, $T_\text{c}$, outer boundary radius, $R_\text{p}$, and hydrogen mass fraction within the interior, $X_{\text{H}_2 \text{, int}}$. We first integrate outwards from the planet's centre, using interior material properties (Section \ref{sec:env_prop}), until the temperature and pressure fall below that of the binodal surface, indicating that a phase transition has occurred and we have transitioned from interior to envelope. We store the mass, radius, pressure and temperature at the binodal surface, labelled as $M_\text{b}$, $R_\text{b}$, $P_\text{b}$ and $T_\text{b}$, respectively. We then integrate outwards again from the binodal, now using the material properties of the envelope (Section \ref{sec:env_prop}). In this region, $X_{\text{H}_2}^\text{gas}$ is increasing as more silicate-rich melt rains out, distilling the envelope to become hydrogen rich. In shooting methods, it is computationally beneficial to shoot from the inner and outer boundary conditions to some intermediate boundary \citep[e.g.][]{kippenhahn2012stellar}. Therefore, we terminate the outward integration when the temperature drops below $1500$~K in the envelope, since at this point $X_{\text{H}_2}^\text{gas}$ has plateaued to a constant value (very close to unity).\footnote{Note that the positioning of this intermediate boundary is chosen purely to aid in numerical convergence, but does not affect the solution found for a given set of boundary conditions.} We again store the mass, radius, pressure and temperature at this intermediate boundary, labelled as $M_\text{i}^{\uparrow}$, $R_\text{i}^{\uparrow}$, $P_\text{i}^{\uparrow}$ and $T_\text{i}^{\uparrow}(=1500\text{ K})$, respectively, where upward-facing arrows denote the outward radial direction of this integration. Finally, we now integrate inwards from the planet's outer boundary. The outer boundary condition is defined with $m=M_\text{p}$, $r=R_\text{p}$, $P=0.1$~bar and $T=T_\text{eq}$. We integrate inwards, terminating at $m=M_\text{i}^{\uparrow}$. Again, we store radius, pressure and temperature at this interface as $R_\text{i}^{\downarrow}$, $P_\text{i}^{\downarrow}$ and $T_\text{i}^{\downarrow}$, respectively. Inevitably, our initial guess of $P_\text{c}$, $T_\text{c}$, $R_\text{p}$  and $X_{\text{H}_2 \text{, int}}$ will mean that radius, pressure and temperature do not match at the intermediate boundary. Hence, we iterate this entire procedure with different combinations of $P_\text{c}$, $T_\text{c}$, $R_\text{p}$, and $X_{\text{H}_2 \text{, int}}$ until we have satisfied the following set of equations:
\begin{equation} \label{eq:matchR}
    \bigg \{ R_\text{i}^{\downarrow} = R_\text{i}^{\uparrow}; \; P_\text{i}^{\downarrow} = P_\text{i}^{\uparrow}; \; T_\text{i}^{\downarrow} = T_\text{i}^{\uparrow} \; (=1500\text{ K}) \bigg \}
\end{equation}
Finally, to close the set of equations, we also require the total integrated hydrogen mass fraction to be equal to that specified in the model:
\begin{equation} \label{eq:matchH2}
    \frac{1}{M_\text{p}} \int_0^{m=M_\text{p}} X_{\text{H}_2}'(m) \; dm = X_{\text{H}_2},
\end{equation}
where $X_{\text{H}_2}'(m)$ is the hydrogen mass fraction as a function of planet mass calculated within a specific iteration of the shooting method. In practice, Equations \ref{eq:matchR}-\ref{eq:matchH2} are solved using the \verb|MINPACK|'s \verb|HYBRD| algorithm with a numerical tolerance of $10^{-6}$. Once a converged model is attained, we post-process the pressure-temperature profile to locate the planet's photospheric radius, defined as the location at which $P=2 g / 3 \kappa$, where $g$ is the gravitational acceleration.

\section{Results} \label{sec:results}

\begin{figure*}
    \includegraphics[width=2.0\columnwidth]{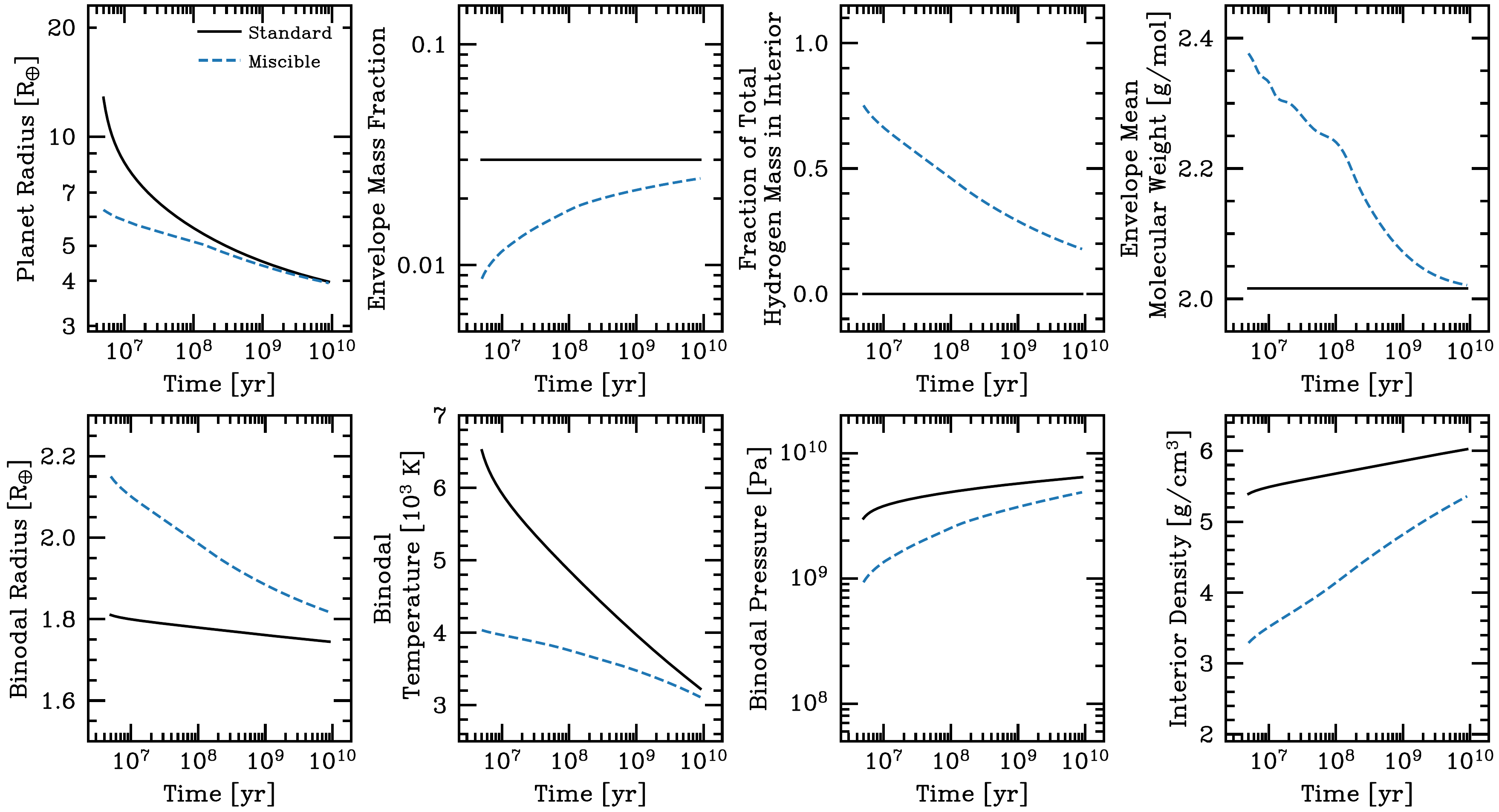}
    \centering
        \cprotect\caption{The evolution of a $6 M_\oplus$ sub-Neptune with equilibrium temperature of $1000$~K and global hydrogen mass fraction of $3$\% for two model classes. A \textsc{standard} model, shown in black, represents a simple case in which silicate melt and hydrogen cannot become miscible. The interior is a pure-silicate sphere of fixed mass. Similarly, the envelope is pure hydrogen and fixed in mass fraction at $3\%$. A \textsc{miscible} model, shown in blue, allows for miscibility between silicate and hydrogen. The interior-envelope boundary is now defined by a binodal. Upper panels show the evolution of each planet's photospheric radius, envelope mass fraction, fraction of total hydrogen mass stored in the planet's interior, and mass-averaged envelope mean molecular weight. Lower panels show the binodal surface in radius, temperature and pressure-space, as well as the interior density. Note that the binodal surface does not exist for the \textsc{standard} model, as it assumes an inert core, hence we show the interior-envelope boundary.} \label{fig:ModelComplexityEvolution} 
\end{figure*}

Our results make comparison between two classes of models to demonstrate the effects of miscibility in sub-Neptunes:
\begin{enumerate}
    \item \underline{\textsc{Standard}:} The simplest conceivable model in which a pure molten silicate interior has a fixed mass. A pure hydrogen envelope is immiscible with the silicate interior. No chemistry is active in this model.
    \vspace{0.5cm}
    \item \underline{\textsc{Miscible}:} Hydrogen is now miscible with the silicate melt within the interior. The binodal surface defines the boundary between miscible interior and immiscible envelope. Silicate vapour and rain are present in the envelope and convection inhibition occurs in the presence of sufficiently steep mean molecular weight gradients.
\end{enumerate}

As a reminder, the sub-Neptunes in our models only consider the H$_2$-MgSiO$_3$ system. Among many additional chemical components, we do not include iron or helium in this initial study. The presence of either would act to increase the density of our modelled planets.

Our main results are summarised in Figure \ref{fig:SchematicEvolution}, showing a schematic for the evolution of these two models. In the \textsc{standard} sub-Neptune model, the interior-envelope boundary is defined as the size of a pure-silicate sphere of a given mass. Over time, the planet contracts as it radiates energy into space, however, the size of the interior remains approximately constant. In this model, the temperature at the boundary can be of order $\sim 10,000$~K when the planet is young, and then cools over time to $\sim 3000$~K after ~Gyrs of evolution. In the \textsc{miscible} model, on the other hand, hydrogen is stored in the envelope \textit{and} interior. The interior-envelope boundary is defined by a binodal surface, representing a phase change within the planet's radial structure. The interior contains a supercritical miscible mixture of hydrogen and silicate. The envelope contains an immiscible mixture of hydrogen and silicate, in both gas and melt phases. The temperature at the binodal surface does not change significantly with time, from $\sim 4000$~K to $\sim 3000$~K over $\sim$~Gyrs of evolution. Its radial position, however, contracts with time and the interior exsolves a larger fraction of hydrogen into the envelope.

An example structure profile for a $6 M_\oplus$ sub-Neptune with equilibrium temperature of $1000$~K and global hydrogen mass fraction of $5$~\% is shown in Figure \ref{fig:3D} as function of pressure, temperature and hydrogen mole fraction. The blue surface represents the binodal. The centre of the planet has a pressure and temperature of $\sim 9000$~K and $\sim 3 \times 10^{11}$~GPa ($\sim 3 \times 10^{6}$~bar). The interior has temperatures and pressures above the binodal surface, meaning that the silicate and hydrogen form a miscible mixture. The hydrogen mole fraction is constant in the interior ($\sim 0.7$, equating to a interior hydrogen mass fraction of $X_{\text{H}_2 \text{, int}} \sim 0.03$). Moving radially out, the temperature and pressure drop until the profile makes contact with the binodal surface. At this point, hydrogen and silicate are immiscible and speciate into a melt phase (silicate-rich rain with dissolved hydrogen), and a gas phase (hydrogen with silicate vapour: SiO, Mg and O$_2$). The mole fractions for these phases are represented as individual black lines on the binodal surface. As temperature and pressure continue to drop as one moves higher in the envelope towards the photosphere of the planet, the silicate-rich melt rains out, meaning that that the mole fraction of hydrogen in the melt phase reduces, while the hydrogen mole fraction of the gas phase increases. This distillation process results in near-pure hydrogen gas at the top of the envelope. Figure \ref{fig:3D} also shows projections of this planetary profile in the pressure-temperature plane (yellow) and hydrogen mole fraction-temperature plane (pink).

\subsection{Evolving models} \label{sec:evolution}
We evolve our planetary models by solving for the conservation of energy:
\begin{equation} \label{eq:evolve}
    \frac{d E_\text{p}}{d t} = - L_\text{rcb},
\end{equation}
where $L_\text{rcb}$ is the radiative luminosity at the outer-most radiative-convective boundary of the model (see Equation \ref{eq:rad_diff}) and $E_\text{p}$ is the total energy of the planet, given by the sum of thermal and gravitational potential energies:
\begin{equation}
    E_\text{p} = \int_0^{M_\text{p}} \bigg( \bigg[\sum_i c_{\text{p},i} \,X_i(m) \,T(m) \bigg]- \frac{G\,m(<r)}{r} \bigg) \;dm,
\end{equation}
where $c_{\text{p},i}$ is the specific isobaric heat capacity for component $i$, including hydrogen and silicate in melt and vapour phases, and $X_i(m)$ is the mass fraction of this component. Heat capacities for all gaseous species, e.g. H$_2$, SiO, Mg and O$_2$, are assumed to be ideal. As discussed in Section \ref{sec:planetary_structure}, our structure models are assumed to be in quasi-equilibrium in that we are implicitly assuming that each snapshot of a planet's history is very well approximated with a stationary model \citep[e.g.][]{Hubbard1977,Fortney2004,Marleau2014}. In practice, we construct a grid of models for a planet of a given mass, hydrogen mass fraction and equilibrium temperature by varying the luminosity at the radiative-convective boundary, $L_\text{rcb}$, from $10^{20}-10^{26}$~erg/s. We then evolve the model by interpolating over this grid and solving for energy conservation (Equation \ref{eq:evolve}) using a $1^\text{st}$-order Euler method and ensuring a sufficiently small timestep $\Delta t \ll |E_\text{p}| / L_\text{rcb}$, where $|E_\text{p}| / L_\text{rcb}$ is the cooling timescale of the planet.

The initial conditions of the models will depend on a planet's accretion history, and atmospheric escape history during protoplanetary disc dispersal, referred to as ``boil-off'' \citep[e.g.][]{Owen2016,Ginzburg2016,Rogers2024a,Tang2024}. Whilst modelling these processes is beyond the scope of this initial study, we choose to initialise all models with the same cooling timescale, which is analogous to the Kelvin-Helmholtz timescale and a useful proxy for planet entropy. We choose a value of $100$~Myrs, motivated by simulations of boil-off \citep[see][]{Owen2017,Rogers2024a}. As a reminder, the cooling timescale is defined as $t_\text{cool} = |E_\text{p}| / L_\text{rcb}$. We discuss our choice in initial conditions in Section \ref{sec:escapeAndAccretion}.

Each model is evolved from $5$~Myrs - $10$~Gyrs. For this initial study, we only consider the thermal evolution of sub-Neptunes, and do not include the effects of atmospheric escape which would act to reduce the hydrogen mass fraction with time.

In Figure \ref{fig:ModelComplexityEvolution}, we show the evolutionary sequence of our fiducial sub-Neptune of $6M_\oplus$ and hydrogen mass fraction of $3\%$ under the \textsc{standard} and \textsc{miscible} model frameworks shown in black and blue, respectively. The upper panels show the planetary radius, envelope mass fraction, fraction of total hydrogen mass stored in the interior and mass-averaged mean molecular weight of the envelope. The lower panels show the position of the binodal surface in radius, temperature and pressure space, as well as the density of the interior. Note that in the \textsc{standard} sub-Neptune model (black), a binodal surface does not exist. Hence, we show the position of the interior-envelope boundary in radius, temperature and pressure space.

In the \textsc{miscible} model, hydrogen is stored in the interior and envelope. As the planet cools, it exsolves a more massive envelope. One can see the fraction of total hydrogen mass stored in the interior reduces from $\sim 75-10\%$. The presence of this hydrogen in the interior reduces its density when compared to the pure-silicate \textsc{standard} interior. When the planet is young, the binodal radius is significantly larger than the \textsc{standard} interior-envelope boundary, meaning the envelope sits further out in the planet's gravitational potential well. As a reminder, the binodal surface represents a location of a phase change within the planet. As it evolves, the binodal temperature reduces marginally from $\sim 4000-3000$~K. The radial position of this surface must contract, which can be understood simply since the binodal surface approximately follows an isotherm with time, as dictated by the binary phase diagram from Section \ref{sec:phase_equilibria}. An isotherm must move inwards for a cooling spherical body, hence the same for the binodal surface. This is in contrast to the \textsc{standard} interior-envelope boundary temperature which cools from $\sim 6500-3000$~K while remaining at an approximately constant size. As the binodal surface temperature drops in the \textsc{miscible} model, less silicate vapour is present in the envelope and so the mass-averaged mean molecular weight also drops with time. 

\begin{figure*}
	\includegraphics[width=2.0\columnwidth]{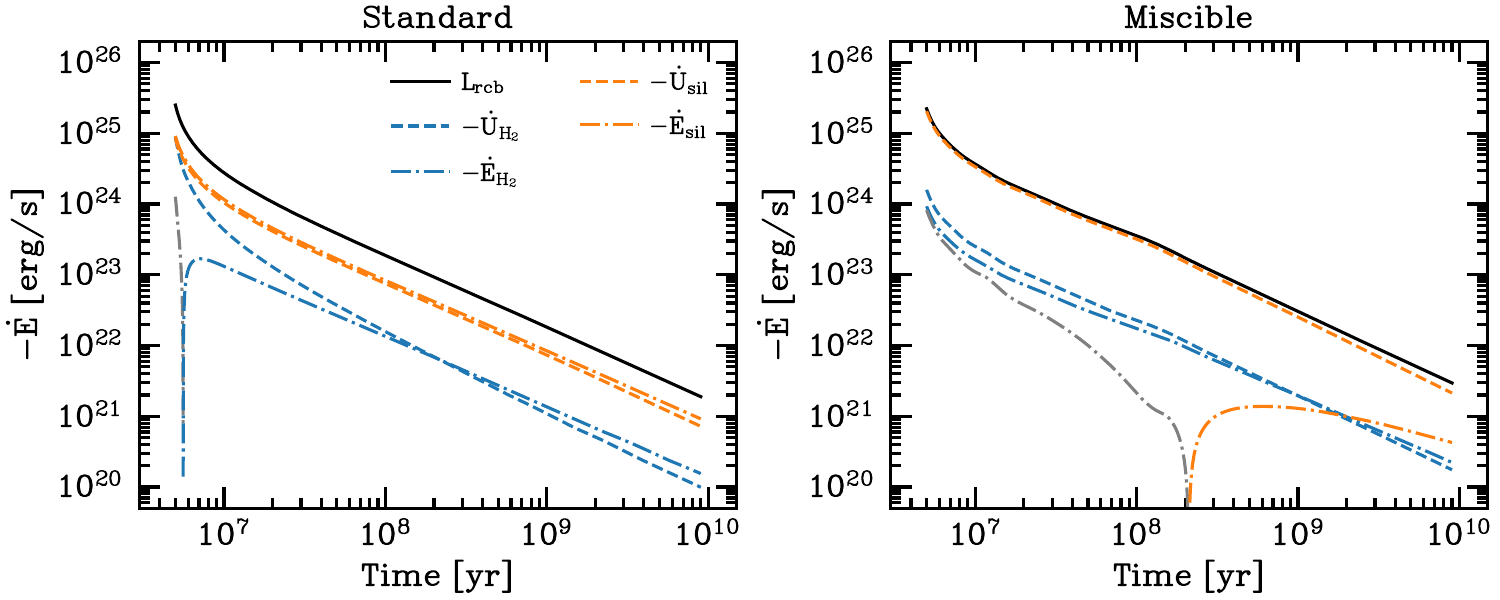}
    \centering
        \cprotect\caption{The energy budget for a $6 M_\oplus$ sub-Neptune with equilibrium temperature of $1000$~K and global hydrogen mass fraction of $3$\%. \textsc{Standard} and \textsc{miscible} models are shown in the left and right-hand panels, respectively. The radiative luminosity at the upper-most radiative-convective boundary is shown in black, and represents the total energy lost from the system. The contributions to this loss in energy come from changes in gravitational binding energy, $-\dot{U}$, shown in dashed lines, and thermal cooling, $-\dot{E}$, shown in dot-dashed lines. These are shown for the silicate and hydrogen mass reservoirs in orange and blue, respectively. Grey lines indicate when an energy loss term becomes negative, implying an increase in a specific energy component.} \label{fig:EnergyBudget} 
\end{figure*}

The gravitational binding energy of a \textsc{miscible} model is necessarily larger (less negative) than that of a \textsc{standard} model of the same radius, since more silicate mass resides higher in the potential well. As a result, for our models which are initialised with the same cooling timescale, equivalent to entropy, \textsc{miscible} models begin more contracted than \textsc{standard} models. Figure \ref{fig:ModelComplexityEvolution} demonstrates that the subsequent radial contraction of a \textsc{miscible} sub-Neptune is slower than that of the \textsc{standard} case. To understand these differences, Figure \ref{fig:EnergyBudget} shows the energy budgets of both models in the left and right-hand side, respectively. The black line shows the radiative luminosity leaving the planet at the uppermost radiative-convective boundary. Since we are not considering any mass loss, this represents all the energy leaving the system at a given time. The orange and blue lines show how this energy is extracted from the silicate and hydrogen mass reservoirs, respectively, with both co-existing in the interior and envelope. Energy can be lost from changes in gravitational binding energy e.g. contraction, $\dot{U}$, shown in dashed lines, or thermal cooling, $\dot{E}$, shown in dot-dashed lines. The sum of the energy extracted through gravitational contraction and thermal cooling for silicate and hydrogen sum to the radiative luminosity at the radiative-convective boundary. If an energy loss term ever turns negative, implying an increase in energy, we represent this with a grey line.

In the \textsc{standard} case, Figure \ref{fig:ModelComplexityEvolution} shows that energy is extracted in comparable amounts through thermal cooling and contraction. Since the silicate dominates the mass of the planet, there is more energy lost from the silicate when compared to the hydrogen. One can see from Figure \ref{fig:ModelComplexityEvolution} that there is a very small contraction of the silicate interior due to the temperature dependence of the silicate melt equation of state. This effect is significant enough to cause a change in the binding energy of the silicate. Coupled with the fact that the \textsc{standard} sub-Neptune model has the maximum envelope mass, since no hydrogen is stored in the interior, these effects combine to cause significant contraction of the planet with time. Note that the thermal energy of the hydrogen, $\dot{E}_{\text{H}_2}$, increases for a very brief phase at the start of the evolution. This is due to rapid contraction causing a small heating effect. This energy source is several factors smaller than the other terms and thus has little effect on the planet's evolution.

In the \textsc{miscible} case, on the other hand, the vast majority of energy is lost from gravitational contraction of silicate. This is because the interior is less dense due to the presence of hydrogen, and so more silicate material sits higher in the gravitational well (see Figure \ref{fig:ModelComplexityEvolution}). Again, the silicate dominates the mass of the planet, and so in order for the planet to contract, the silicates must settle deeper into the gravitational well, which dominates the energy budget. While the silicate is settling deeper into the gravitational potential, hydrogen exsolves from the interior. As a result, the planet does not contract as rapidly as the in the \textsc{standard} sub-Neptune model. The next dominant sources of energy loss are the gravitational contraction and cooling of the of the hydrogen. Lastly, the smallest contribution to changes in energy is the thermal cooling of the silicate. In fact, for the first $\sim 200$~Myrs, the silicate's thermal energy increases. This is because the central temperature of the planet also gradually increases with time since the binodal surface increases in pressure for a similar, although also gradually decreasing interior mass. This compression of the interior causes the central temperature to increase. This is also in contrast to the \textsc{standard} model, for which the central temperature is monotonically decreasing with time.

\subsection{Varying hydrogen mass fraction and planet mass} \label{sec:Mp_XH2}

\begin{figure*}
	\includegraphics[width=2.0\columnwidth]{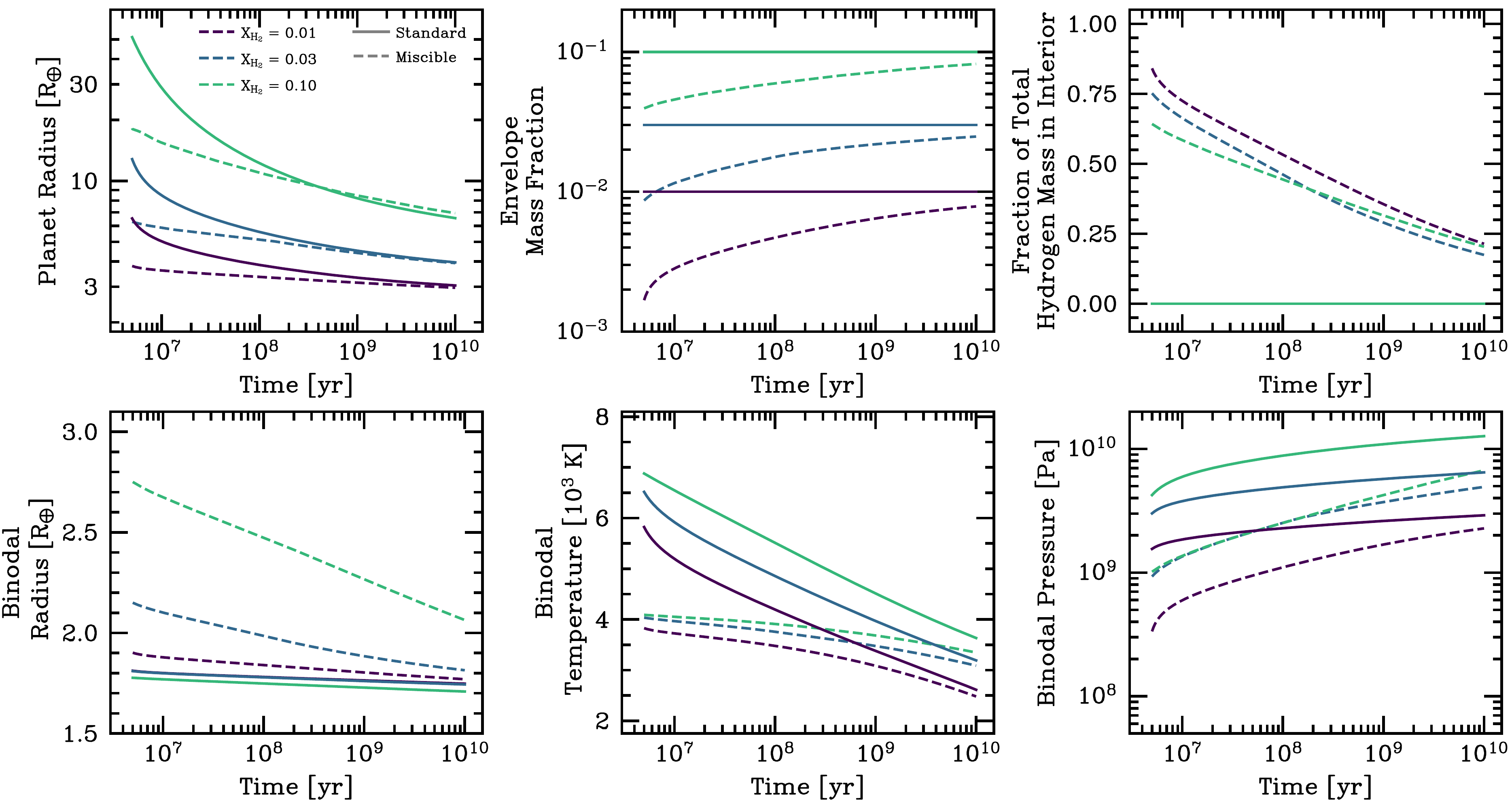}
    \centering
        \cprotect\caption{The evolution of $6 M_\oplus$ sub-Neptunes with equilibrium temperature of $1000$~K and global hydrogen mass fractions of $1$\%, $3\%$ and $10\%$ for two model classes. A \textsc{standard} model, shown in solid lines, represents a simple case in which silicate melt and hydrogen cannot become miscible. The interior is a pure-silicate sphere of fixed mass. Similarly, the envelope is pure hydrogen and fixed in mass fraction. A \textsc{miscible} model, shown in dashed lines, allows for miscibility between silicate and hydrogen. The interior-envelope boundary is now defined by a binodal. Upper panels show the evolution each planet's photospheric radius, envelope mass fraction, and fraction of total hydrogen mass stored in the planet's interior. Lower panels show the binodal surface in radius, temperature and pressure-space. Note that the binodal does not exist for the \textsc{standard} model, hence we show the interior-envelope boundary.} \label{fig:fH2Evolution} 
\end{figure*}

We now investigate the effects of varying the hydrogen mass fraction and planet masses of sub-Neptunes. In Figure \ref{fig:fH2Evolution}, we show the evolution of $6M_\oplus$ sub-Neptune models with hydrogen mass fractions of $1\%$, $3\%$, and $10\%$. \textsc{Standard} and \textsc{miscible} models are shown in solid and dashed lines, respectively. General results are shared for both classes of models. Sub-Neptunes with larger hydrogen mass fractions are larger in size with interior-envelope boundaries (defined by the binodal surface in the \textsc{miscible} case) existing at larger radii, temperatures and pressures. However, there is very little variation in the binodal temperature in \textsc{miscible} model for varying hydrogen mass fraction. Again, this is because the binodal surface is dictated by the phase diagram of the H$_2$-MgSiO$_3$ binary system.

\begin{figure*}
	\includegraphics[width=2.0\columnwidth]{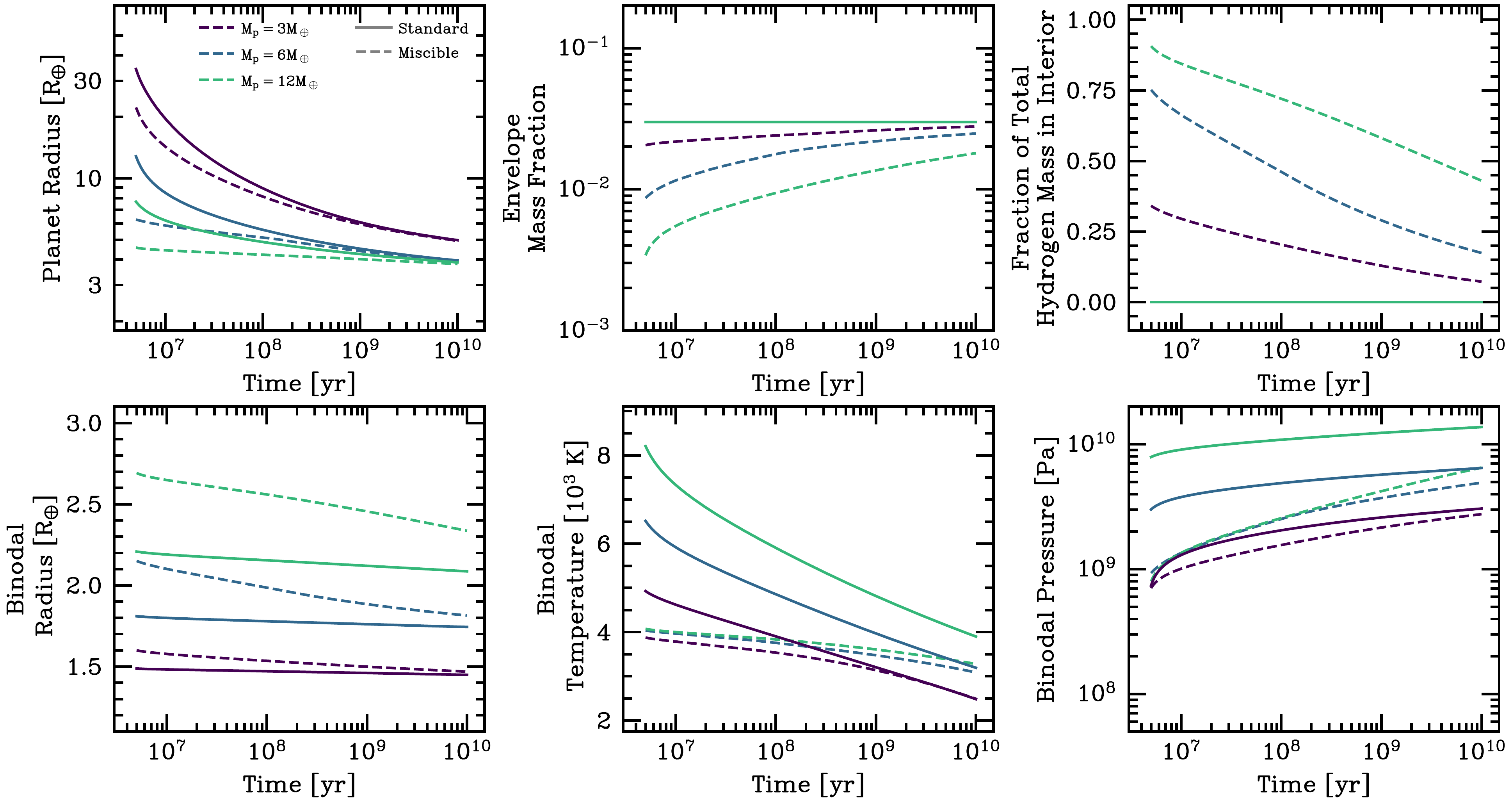}
    \centering
        \cprotect\caption{Same as Figure \ref{fig:fH2Evolution} but for sub-Neptunes of varying masses, including $3M_\oplus$, $6M_\oplus$ and $12M_\oplus$. All planets have a hydrogen mass fraction of $3\%$ and equilibrium temperature of $1000$~K. \textsc{Standard} and \textsc{miscible} models are shown in solid and dashed lines, respectively.} \label{fig:MpEvolution} 
\end{figure*}

In Figure \ref{fig:MpEvolution}, we show the same evolution but now for varying planet masses of $3M_\oplus$, $6M_\oplus$ and $12M_\oplus$ with a hydrogen mass fraction of $3\%$. One can see that smaller mass planets are larger since their gravitational potential wells are shallower and so the hydrogen envelope is less bound. The effect is compounded in the \textsc{miscible} models by the fact that smaller mass planets have lower central temperatures and pressures, and thus store less hydrogen in their interiors. As a result, smaller mass planets host larger envelope mass fractions, further increasing their radii when compared to larger mass planets. We highlight that atmospheric escape, if present, will preferentially remove hydrogen-dominated material from lower mass planets, which would act to reduce their size. We discuss this in Section \ref{sec:escapeAndAccretion}.

\begin{figure*}
	\includegraphics[width=2.0\columnwidth]{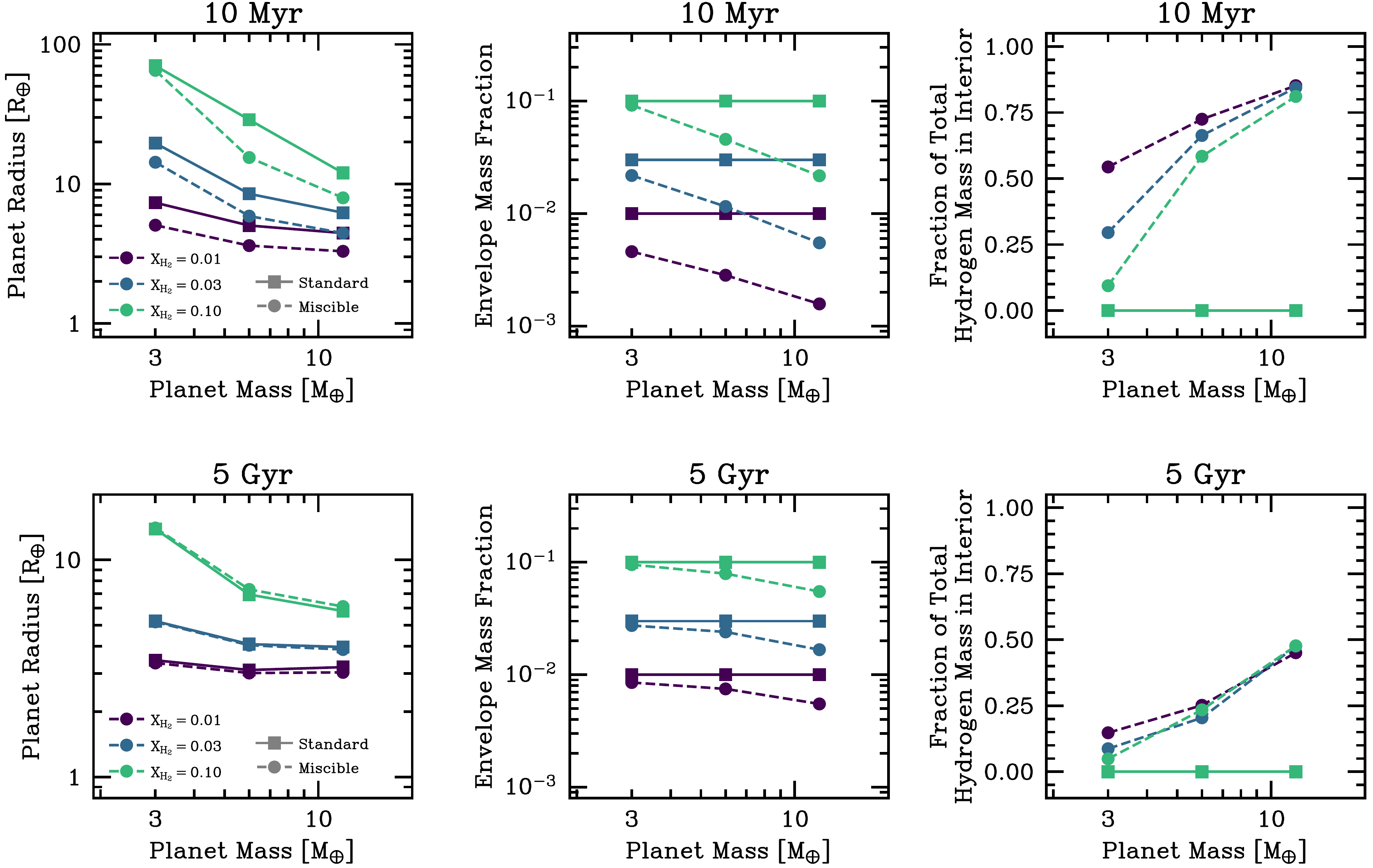}
    \centering
        \cprotect\caption{Planet properties at $10$~Myrs and $5$~Gyrs are shown across a range of planet masses and hydrogen mass fractions in the upper and lower panels, respectively. \textsc{Standard} and \textsc{miscible} models are shown in solid-squares and dashed-circles, respectively. Planet photospheric radii, envelope mass fractions and fractions of total hydrogen mass stored in interiors are shown as a function of planet mass in the left, central and right-hand panels, respectively. Note that planet models only include hydrogen and silicate, without the inclusion of iron or helium. As a result, these models should not be used to compare with exoplanet observations.} \label{fig:MassRadius} 
\end{figure*}

Finally, in Figure \ref{fig:MassRadius}, we show mass-radius relations for our grid of models across planet mass and hydrogen mass fraction. We also show envelope mass fractions and fractions of total hydrogen mass stored in planet interiors. Models are evolved for $10$~Myrs in the upper panels and $5$~Gyrs in the lower panels. These relations should thus be interpreted as planetary isochrones. \textsc{Standard} and \textsc{miscible} models are shown in solid and dashed lines, respectively.

Planet radii differ between \textsc{standard} and \textsc{miscible} models by several tens of percent at young ages, with miscibility resulting in systematically smaller planet sizes. This is more apparent at earlier epochs since planets with higher energy store more of their hydrogen in their interiors, reducing envelope masses and enriching envelopes with more silicate vapour. As discussed in Section \ref{sec:evolution}, this causes silicate to sit higher in the gravitational well and also results in slower contraction. However, there is very little difference between \textsc{standard} and \textsc{miscible} planet properties at $5$~Gyrs. As shown in Figures \ref{fig:ModelComplexityEvolution}, \ref{fig:fH2Evolution} and \ref{fig:MpEvolution}, \textsc{miscible} sub-Neptunes exsolve larger envelopes with time, eventually converging close to the \textsc{standard} envelope mass fraction. In addition, Figure \ref{fig:ModelComplexityEvolution} shows that envelopes become more hydrogen rich with time, meaning the chemical composition of \textsc{standard} and \textsc{miscible} models also converge. The temperature at the interior-envelope boundary additionally converges over Gyr timescales. These effects all contribute to the similarities in planet properties at $5$~Gyrs shown in Figure \ref{fig:MassRadius}. One can see that \textsc{standard} models are fractionally larger in size for a given mass. This is because \textsc{miscible} models have smaller envelope mass fractions, despite their binodal radii being larger than \textsc{standard} interior-envelope boundaries. The only exception is for largest masses and largest hydrogen mass fractions (e.g $12M_\oplus$ with a hydrogen mass fraction of $10\%$), in which case the binodal radius is substantially larger, meaning the \textsc{miscible} model is marginally larger in size than the \textsc{standard} model.

Finally, we highlight again that our interior structure models do not include the presence of iron or helium, among various other chemical components. Iron would predominantly act to increase the density of the interior \citep{Young2025}, while helium would act to increase the density of the envelope. As a result, we would expect planet radii to be systematically smaller (of order $\sim$~tens of per cent) for a given mass with the inclusion of iron and helium. Since both of these chemical components are thought to be prevalent in sub-Neptunes, the mass-radius relations shown in Figure \ref{fig:MassRadius} should not be used to compare with exoplanet observations. 

\section{Discussion} \label{sec:discussion}

Our results suggest the need for a conceptual shift in the way we interpret the interiors of sub-Neptunes. In previous studies, the simplest sub-Neptune models have considered a hydrogen-dominated envelope atop an inert silicate-rich interior. Building from this, some studies have also considered the chemical interactions at the interior-envelope boundary, including the evaporation or dissolution of gases into and out of the silicate magma ocean. In this study, we address the fact that silicate melt and hydrogen are expected to be fully miscible above temperatures of $\sim 4000$~K. We define the \textit{binodal surface} to delineate regions in which silicate and hydrogen become miscible or immiscible. The binodal surface contracts with time, and the interior exsolves a larger envelope mass as the planet cools. Here we discuss implications, limitations and future work.

\subsection{Miscibility vs. solubility}
Previous works have investigated the dissolution of gases into or out of a magma ocean by means of a solubility framework \citep[e.g.][]{Paonita2005,Chachan2018,Kite2020,Kite2021,Lichtenberg2021b,Schlichting2022,Charnoz2023,Rogers2024b,Shorttle2024,Nicholls2024,Vazan2024,Werlen2025,Werlen2025b,Ito2025}. Typically, this is done with Henry's law, which itself is an approximation to the full non-ideal mixing framework described in Section \ref{sec:phase_equilibria} in the limit of low temperatures and low solute concentrations \citep[e.g][]{White_2020}. Consider Henry's law applied to hydrogen gas dissolving into a silicate melt:
\begin{equation}
    x_{\text{H}_2}^{\text{melt}} = h_\text{c} \,f_{\text{H}_2},
\end{equation}
where $x_{\text{H}_2}^{\text{melt}}$ is the hydrogen mole fraction dissolved in the silicate magma ocean, $h_\text{c}$ is the appropriate Henry's law constant that is independent of concentration and often determined from experiments, and $f_{\text{H}_2}$ is the hydrogen fugacity in the atmosphere (equivalent to partial pressure in an ideal system). This relation is a good approximation to the full non-ideal mixing framework in the limit of low temperatures and low H$_2$ concentrations. In order to capture the mutual solubility of silicate and H$_2$ with this approach, as evidenced by the binodal, analogous equations would also be needed for the hydrogen-rich phase, simulating silicate evaporation into a supercritical hydrogen envelope with appropriate Henry's law constants. However, considering the full binodal surface, one finds a strong divergence from Henry's law at intermediate mole fractions, such as near the crest of the binodal, which is relevant for the interiors of sub-Neptunes (see Figure \ref{fig:3D}). If Henry's law is extrapolated to arbitrary temperatures and larger solute mole fractions (for a given pressure), then one would miss the existence of a phase change as the two chemical components become entirely miscible. Previous studies have thus likely over/underestimated the gas content dissolved in magma oceans in chemical equilibrium. This highlights the impact of miscibility on the geochemistry occurring in sub-Neptunes, in addition to the physical structure and thermal evolution discussed in this paper.

\subsection{On the prevalence of convection inhibition} \label{sec:convectionInhib}

\begin{figure*}
	\includegraphics[width=2.0\columnwidth]{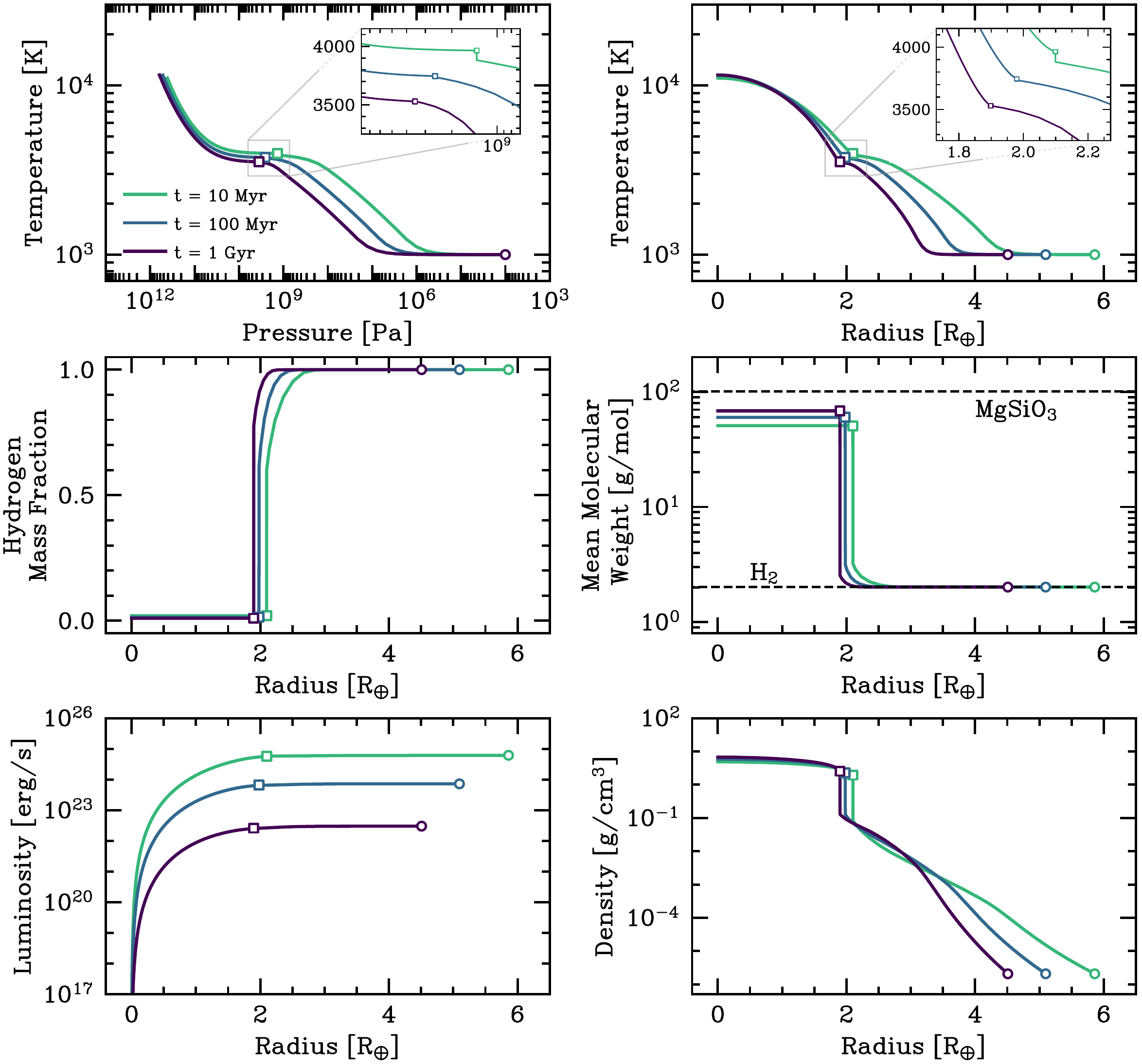}
    \centering
        \cprotect\caption{The interior structure profile for a $6 M_\oplus$ sub-Neptune with equilibrium temperature of $1000$~K and global hydrogen mass fraction of $3$\% at $10$~Myr, $100$~Myr and $1$~Gyr. Binodal surfaces are shown as squares, whilst photospheric radii are shown as circles. Upper panels show pressure-temperature and radius-temperature profiles, middle panels show the hydrogen mass fraction and mean molecular weight as a function of radius, and lower panels show luminosity and density profiles. Insets in the upper panels demonstrate the diminishing convection inhibition near the binodal surface as the planet cools over time. Note that luminosity profiles are calculated by post-processing evolution models.}
        \label{fig:ProfileEvolution} 
\end{figure*}

Multiple studies have shown that mean molecular weight gradients can inhibit convection in the deep interior of sub-Neptune envelopes \citep[e.g.][]{Leconte2017,Ormel2021,Misener2022,Steinmeyer2024,Leconte2024,Vazan2024}. Whereas modelling efforts have considered the evaporation, dissolution, and rain-out of magma ocean chemical components, our work introduces a new framework which revolves around phase equilibria. We do not define a magma ocean surface, but instead a binodal surface, which separates the miscible H$_2$-MgSiO$_3$ interior from the immiscible envelope with separate gas and melt phases. We use the phase diagram for the binary H$_2$-MgSiO$_3$ system to self-consistently calculate the mole fractions of each component in each phase. We do not, however, allow for chemical reactions to take place between these chemical components to produce other species, such as water \citep{Misener2023}.

Figure \ref{fig:ProfileEvolution} shows evolution profiles of our fiducial $6M_\oplus$ sub-Neptune with a hydrogen mass fraction of $3\%$ and equilibrium temperature of $1000$~K at $10$~Myr, $100$~Myr and $1$~Gyr. We show the pressure-temperature and radius-temperature profiles in the upper panels and hydrogen mass fraction and mean molecular weights profiles in the middle panels. 

Convection inhibition occurs when there is a sufficiently steep mean molecular weight gradient caused by the presence of condensible species (SiO, Mg and O$_2$, in our case) that are heavier than the background gas (hydrogen, in our case). By necessity, this can only occur just above the binodal surface, since the reservoir for silicate vapour species is in the interior. Convection inhibition is unlikely to occur in the miscible interior since miscible mixtures can mix very efficiently, and the interior is expected to be fully molten and vigorously convecting. The insets in Figure \ref{fig:ProfileEvolution} show that a very small convection-inhibited region exists at the earliest evolutionary phases. Since heat is transported predominantly through radiative diffusion and conduction, the temperature gradient is much steeper than that of a convective moist adiabat. As a result, the temperature drops by $\sim 100$~K through an increase in altitude of $\sim 100$~m. The younger the planet, the hotter the binodal surface, and thus the more silicate vapour species are present in the envelope. This is confirmed by lower hydrogen mass fractions and higher mean molecular weights as a function of radius. After $\sim 100$~Myrs, the convection-inhibited zone is gone. We confirm that models in which convection inhibition is not permitted are nearly indistinguishable from our \textsc{miscible} models, meaning this mechanism has negligible effects on a planet's long-term evolution. 

We note that previous studies have found a more significant impact of convection inhibition on planetary evolution. \citet{Vazan2024} considered the effects of convection inhibition in the context of pebble rain-out in the post-formation phase. As a result, the silicate mass in the envelope was significantly larger and settled from further out in the gravitational well than in our case. This resulted in more substantial convection-inhibited zones than seen in our work. This latter study, along with other previous investigations of convection inhibition \citep[e.g.][]{Misener2022,Misener2023}, have not included the effects of miscibility. As a result, temperatures at interior-envelope boundaries were higher, introducing more silicate vapour in the envelope and thus larger regions for which steep mean molecular weight gradients can inhibit convection. In addition, explicitly including the chemical production of water in the envelope, results in more silicate vapour in the envelope than when only the silicate-hydrogen system is considered \citep{Misener2023}.

In regions of a planet in which convection is not the dominant energy transport mechanism, one requires the internal luminosity passing through the region to calculate the temperature gradient (Equation \ref{eq:rad_diff}). In our model, we have assumed a constant luminosity throughout the planet, which comes from the assumption of models in quasi-equilibrium, e.g. $\partial / \partial t = 0$ in the energy conservation equation:
\begin{equation} \label{eq:dLdm}
    \frac{\partial L}{\partial m} = -c_\text{p} \frac{\partial T}{\partial t}  + \frac{\delta}{\rho}\frac{\partial P}{\partial t},
\end{equation}
where the first and second terms represent energy released due to thermal cooling and $PdV$ work, respectively. To validate our assumption of constant luminosity, we post process our fiducial $6M_\oplus$ model by integrating Equation \ref{eq:dLdm} through the planet from a lower boundary condition of $L(m=0) = 0$ at $10$~Myr, $100$~Myr and $1$~Gyr. Time derivatives in Equation \ref{eq:dLdm} are calculated using a first-order Euler method between successive profiles in the evolutionary sequence. These luminosity profiles are shown in the bottom panel of Figure \ref{fig:ProfileEvolution}. One can see that the luminosity is very close to being constant above the binodal surface. In all cases, we find a fractional increase in luminosity of $<10\%$ between the binodal surface and photosphere. We also confirm that the luminosity at the uppermost radiative-convective boundary is within $10\%$ of the value asserted for each model as a free parameter (see Section \ref{sec:numerical_proc}). This implies that Equation \ref{eq:dLdm} is only weakly coupled to the remaining structure Equations \ref{eq:mass_cons}, \ref{eq:hydro_eq} and \ref{eq:heat_transport} in this physical problem. We note that the luminosity is not required to calculate the temperature gradient in convective regions under the assumption of an adiabatic temperature gradient, such as the interior of the planet where luminosity is clearly not constant from Figure \ref{fig:ProfileEvolution}. This validates our assumption of a constant luminosity in calculating temperature gradients in convection-inhibited regions, particularly noting the significant uncertainties in radiative and conductive opacities in this region also required to calculate temperature gradients in Equation \ref{eq:rad_diff}.

% However, In non-static models, the radiative luminosity at the interior-envelope boundary can be many orders of magnitude lower than that at the outer-most radiative-convective boundary \citep[see Figure 8 of][]{Owen2016}. This means that the radiative diffusion temperature gradient would also be many orders of magnitude lower at this location, meaning the steep temperature gradient in convection-inhibited zones would be less extreme.}} On the other hand, we highlighted in Section \ref{sec:env_prop} that gas opacities may be significantly higher than that of Solar metallicity gas in the deep envelope due to silicate vapour species. Since the radiative diffusion temperature gradient (Equation \ref{eq:rad_diff}) is also proportional to the opacity, this would act to steepen the temperature gradient in convection-inhibited zones. However, we also confirm the findings of \citet{Misener2022,Misener2023}, in that conduction dominates the heat transport at the relevant temperatures and pressures, e.g. $\kappa_\text{c} \gg \kappa$ in Equation \ref{eq:kappa_eff}, meaning that gas opacities would need to increase by \textit{several} orders of magnitude before its effect would be marginally noticed in the structure and evolution of the planet. 

\subsection{The effects of atmospheric accretion and escape} \label{sec:escapeAndAccretion}
Our study has only considered the thermal evolution of sub-Neptunes, and ignored atmospheric escape and gas accretion, which will certainly affect the evolution of such highly-irradiated planets. 

The planet formation sequence, including solid and gas accretion, will be strongly affected by the miscibility between hydrogen and silicate. As a progenitor planetary embryo forms, it must remain molten in order for the miscibility with hydrogen to take effect. In this study, we have implicitly assumed that the entire interior is molten and thus available to mix with hydrogen. If this is not the case, then only a fraction of the planet's interior will be miscible with the accreting hydrogen gas. One way to interpret this scenario is a hybrid between our \textsc{standard} and \textsc{miscible} models, since \textsc{standard} models do not allow any inclusion of hydrogen in the interior. 

Nevertheless, assuming some significant portion of the interior is molten and available for miscibility with the envelope, then the gas accretion process will likely be altered by the consequences of hydrogen-silicate miscibility. In a standard model for sub-Neptune structure, gas accretion requires the envelope to contract deeper within the Bondi radius such that more gas becomes graviationally bound onto the planet from the local protoplanetary disc \citep[``to cool is to accrete'', e.g.][]{Lee2015,Ginzburg2016}. However, miscibility will allow most of this hydrogen to be stored in the interior. It is unclear how the competition between miscibility and accretion will play out during the disc phase, hence this investigation is left for future work. This also extends to atmospheric escape during protoplanetary disc dispersal, referred to as atmospheric ``boil-off'' \citep{Owen2016,Ginzburg2016,Rogers2024a,Tang2024}. In the standard model, the dramatic loss of envelope material is expected to prematurely cool a planet, which may expedite the envelope exsolution process from the interior. It also sets the initial entropy of the planet for the post-disc phase. While we have accounted for this by setting all planets to have the same initial cooling timescale of $100$~Myrs (which in turn sets the planet entropy), motivated by boil-off simulations from \citet{Owen2016,Rogers2024a}, the initial cooling timescale may be different if miscibility is allowed to act. We leave this for future work.

In the post-disc phase, stellar-driven atmospheric escape will continue to sculpt sub-Neptune properties \citep[e.g.][]{Owen2013,LopezFortney2013}. The mass-radius relations at $5$~Gyrs presented in Figure \ref{fig:MassRadius} will certainly be affected by this process. Lower mass, highly irradiated planets will lose more atmosphere, which acts to produce a positive gradient in the mass-radius relation of sub-Neptunes, as observed in the exoplanet demographics \citep[see][]{Rogers2023b}. For the most irradiated planets, atmospheric escape can strip entire hydrogen-dominated envelopes, producing super-Earths. Under our framework of miscibility, this is still expected to occur. As the envelope is removed over $10-100$~Myrs, the interior will continue to exsolve its hydrogen reservoir until it has cooled below the solidus of a silicate mixture and crystallises. At this point, any remaining hydrogen may become locked in the interior, producing slightly under-dense interiors when compared to Earth \citep[e.g.][]{Rogers2024b, Rogers2025}.

We have stressed that our planet models do not include iron or helium, meaning our planet radii are systematically larger than in models that would include these chemical components. However, the general qualitative results should remain the same. One can see from Figure \ref{fig:MassRadius} that the introduction of interior-envelope miscibility does not significantly alter the size of planets after Gyrs of evolution. However, one can also see that the envelope masses can be several factors smaller, due to significant hydrogen mass fractions stored in the planets' interiors. This means that, while mass-radius inferences may infer the correct hydrogen mass fraction, the envelope mass fraction may be over-estimated by several factors. This is particularly true for more massive planets, which exsolve smaller mass envelopes. Since we also expect larger mass planets to retain larger hydrogen mass fractions due to atmospheric escape, it may be that the envelope mass fractions (but not the hydrogen mass fractions) are relatively constant across a broad range in sub-Neptune masses as a result of these competing effects.

\subsection{Model Limitations}
The goal of our modelling efforts was to re-evaluate the interior structure of sub-Neptunes in light of miscibility of hydrogen and silicate melt \citep{Markham2022,Young2024,Stixrude2025}. However, many assumptions were made to simplify this initial framework. Firstly, and as previously stated, other chemical species such as iron \citep{Young2025} and helium were not included. In addition we have not included H$_2$O, which has recently been shown to be miscible with H$_2$ across a broad range of parameter space \citep[e.g.][]{Gupta2024}. Ideally, DFT simulations are required for the H$_2$-He-H$_2$O-Fe-MgSiO$_3$ system in order to cover a minimally spanning sub-Neptune model with all expected major chemical components. Note that chemical equilibrium can also change the abundance of these components, such as with the production of endogenous water \citep[e.g.][]{Kite2021,Schlichting2022,Young2023,Rogers2024b,Rogers2025,Werlen2025b}. In our model, we assume silicate vapour speciates into gaseous SiO, Mg and O$_2$. However, \citet{Misener2023} showed that these species react with the background hydrogen to produce H$_2$O and SiH$_4$ (silane), which can change the mean molecular weight of the envelope. Whilst this effect will not significantly change the thermal evolution and structure of sub-Neptunes, it will affect the presence of any silicate vapour species in the upper atmosphere, which may be detected with atmospheric spectroscopy.

As always, interior structure modelling is dependent on our knowledge of material equations of state. We have assumed equations of state for silicate melt from \citet{deKoker2009,Wolf2018} and hydrogen from \citet{Chabrier2019}. However, the equations of state for the mixture of these species is currently unknown. We assumed ideal mixing between the two equations of state in the interior, as supported by DFT simulations \citep[e.g.][]{Young2024,Young2025}, however, this does not include changes to adiabatic temperature gradients, thermal expansivities, or heat capacities. Similar to equations of state, our knowledge of opacities and conductivities of hydrogen and silicate mixtures is limited at high pressures and temperatures \citep[][]{Eberlein2025}.

Other assumptions of our model include a fully convective and molten interior with adiabatic temperature gradient, as well as perfect rain-out of silicates from the envelope. \citet{Leconte2017} showed that condensation-inhibited convection can lead to stable radiative or conductive layers near a cloud layer, which our model does not account for. The convection-inhibited zones seen in our models are small and thus inconsequential for the interior structure and evolution of such planets. However, if condensation-induced convection-inhibited zones form, then the assumption of constant luminosity must be reassessed. These model improvements are left for future work. 

\section{Conclusions} \label{sec:conclusion}
In this study, we have introduced a framework for sub-Neptune interior structure and thermal evolution that accounts for the miscibility between H$_2$ and MgSiO$_3$. We compared this model to a standard model consisting of an inert interior and a non-interacting hydrogen envelope in which no miscibility can occur. Our conclusions are as follows:

\begin{itemize}
    \item The binodal surface for the hydrogen-silicate phase diagram represents a boundary between regions of parameter space where hydrogen and silicate are miscible or immiscible. We advocate for using the binodal to demarcate the boundary between interior and envelope within a sub-Neptune. The ``interior'' is defined as the region in which temperatures and pressures are above the binodal surface, meaning hydrogen and silicate form a supercritical, miscible mixture. The ``envelope'' is defined as the region in which temperatures and pressures are below the binodal surface, meaning hydrogen and silicate are immiscible and exist in both gaseous and melt (rain) phases.
    \vspace{0.4cm}
    
    \item Sub-Neptunes store a significant fraction (several tens of per cent) of their hydrogen mass in their interiors as a direct result of miscibility. As a planet thermally cools and its radius contracts, its interior exsolves a larger mass envelope and the binodal surface contracts. Temperatures at the binodal surface reduce from $\sim 4000-3000$~K.
    \vspace{0.4cm}

    \item Since sub-Neptune interiors hold large quantities of hydrogen, their densities are reduced from that of pure silicate. As a result, silicate material sits higher in the gravitational potential well and alters how sub-Neptunes contract. Instead of energy being lost predominantly from thermal cooling \textit{and} contraction of the hydrogen envelope, as is the case in a standard model, the vast majority of energy is lost from a miscible sub-Neptune due to gravitational contraction of the silicate melt, while exsolving some of the hydrogen from the interior. This effect means that miscible sub-Neptunes contract slower than planets in the standard scenario.
    \vspace{0.4cm}

    \item When a sub-Neptune is young ($\sim 10-100$~Myrs), the temperature at the binodal surface is hotter, meaning more hydrogen is stored in the interior and more silicate vapour is present in the envelope. This latter effect means the bulk envelope mean molecular weight is elevated, and mean molecular weight gradients can inhibit convection. These effects combine to cause the radii of a young miscible sub-Neptunes of a given mass and hydrogen mass fraction to be reduced when compared to the case without miscibility.
    \vspace{0.4cm}

    \item After $\sim$~Gyrs of evolution, sub-Neptune interiors have exsolved most of their hydrogen, and the envelope holds less silicate vapour with no convection inhibition. As a result, the radii of sub-Neptunes with or without the effects of miscibility are very similar for a given mass during this epoch. However, the radial distribution of hydrogen and silicates still differs and some hydrogen remains stored in the planet's interior.
    \vspace{0.4cm}

    \item A solubility framework, such as Henry's law, is not sufficient to capture the significant effects of miscibility on sub-Neptune structure and thermal evolution. Crucially, extrapolation of such frameworks to arbitrary temperatures and concentrations (as is required for sub-Neptunes) misses the existence of phase changes within the planet.

\end{itemize}

In this initial study, we have neglected the effects of additional chemical components, including iron, helium and water. Since all of these species are expected to be present in sub-Neptunes, future studies should include them in a self-consistent thermochemical framework. This also requires further work on understanding their phase equilibria and equations of state. Furthermore, future work should be done on the implications of miscibility on planet formation, gas accretion and atmospheric escape, to understand how we may provide robust observational tests as to the prevalence of interior-envelope miscibility within sub-Neptunes.

\section*{Acknowledgements}
We kindly thank the anonymous reviewer for comments that helped improve the paper. JGR gratefully acknowledges support from the Kavli Foundation.  EDY  acknowledges financial support from NASA grant 80NSSC24K0544 (Emerging Worlds program). HES gratefully acknowledges support from NASA grant 80NSSC25K7143 (Exoplanet Research Program).
%%%%%%%%%%%%%%%%%%%%%%%%%%%%%%%%%%%%%%%%%%%%%%%%%%
\section*{Data Availability}

All models will be made available upon reasonable request.



\bibliographystyle{mnras}
\bibliography{references} % if your bibtex file is called example.bib

%\begin{thebibliography}{99}
%\end{thebibliography}

%%%%%%%%%%%%%%%%%%%%%%%%%%%%%%%%%%%%%%%%%%%%%%%%%%

%%%%%%%%%%%%%%%%% APPENDICES %%%%%%%%%%%%%%%%%%%%%

\appendix

\section{An analytic fit to the binodal surface} \label{app:binodal}

\begin{figure}
	\includegraphics[width=\columnwidth]{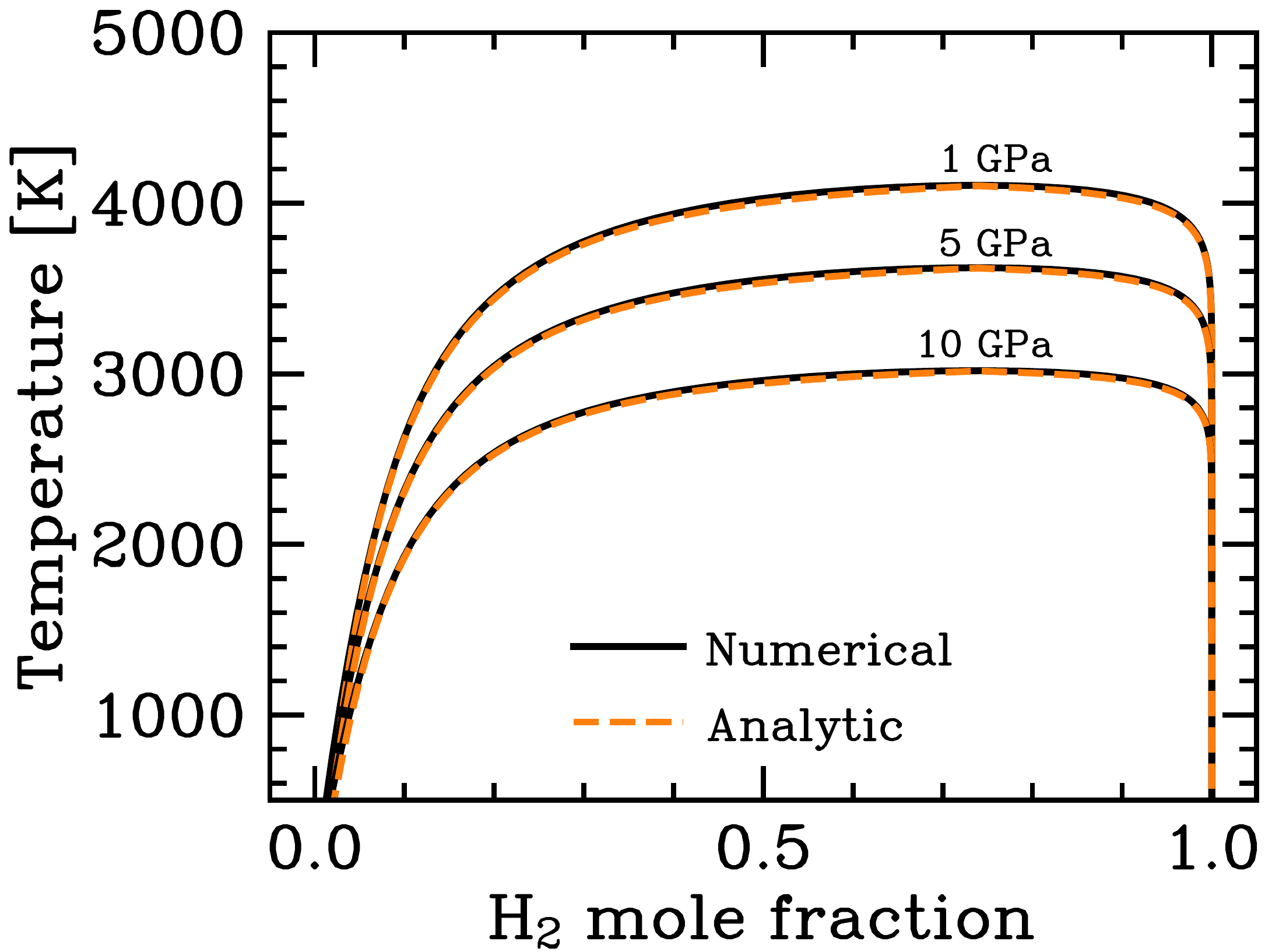}
    \centering
        \cprotect\caption{The binodal surface, as described in Section \ref{sec:phase_equilibria} is shown in black for various pressures. The analytic fit to the this surface, as provided in Equations \ref{eq:analytic_Tb_first}-\ref{eq:analytic_Tb_last}, is shown in orange-dashed lines.}  \label{fig:AnalyticBinodal} 
\end{figure}

Calculating the partitioning of the H$_2$-MgSiO$_3$ system into melt and gas phases from first-principles involves finding the mole fractions of each phase that produce identical chemical potentials with a positive second derivative of the Gibbs free energy (Section \ref{sec:phase_equilibria}). Given the extreme sensitivity of the binodal surface, this is computationally challenging, especially noting that high numerical tolerance is required for converged solutions within the interior structure solver described in Section \ref{sec:numerical_proc}. To mitigate this problem, we pre-calculate the binodal surface to a high accuracy and then fit analytic functions to its form as a function of hydrogen mole fraction, $x_{\text{H}_2}$, and pressure, $P$, in GPa. This is shown in Figure \ref{fig:AnalyticBinodal}. The analytic binodal surface temperature, $T_\text{b}$ is given by a piece-wise function:
\begin{equation} \label{eq:analytic_Tb_first}
  \log_{10}[T_\text{b}(x_{\text{H}_2}, P)]= \left\{
  \begin{array}{lr} 
      f(\tilde{x}) & x_{\text{H}_2} \leq x_\text{c} \\
      g(\tilde{y}) & x_{\text{H}_2} > x_\text{c}
      \end{array}
\right.  
\end{equation}
where $x_\text{c}=0.73913$ is the critical mole fraction at the crest of the binodal surface \citep{Stixrude2025} and $\tilde{x}$ and $\tilde{y}$ are transformed variables, defined by:
\begin{equation}
    x_{\text{H}_2} \rightarrow \tilde{x} = \log_{10} \bigg( \frac{x_{\text{H}_2}}{x_\text{c}} \bigg ) \;\;\;\; x_{\text{H}_2} \leq x_\text{c},
\end{equation}
\begin{equation}
    x_{\text{H}_2} \rightarrow \tilde{y} = \log_{10} \bigg( 1- \frac{x_{\text{H}_2} - x_\text{c}}{1 - x_\text{c}} \bigg ) \;\;\;\; x_{\text{H}_2} > x_\text{c}.
\end{equation}
The ascending function, $f$, is defined as:
\begin{equation}
    f(\tilde{x}) = \alpha_1 + \alpha_2 \bigg[ 1 + \alpha_3 \; \exp \{- \alpha_4 (\tilde{x} - \alpha_5)\} \bigg]^{-\frac{1}{\alpha_3}},
\end{equation}
where:
\begin{equation}
    \alpha_1=\log_{10} T_\text{c} - \alpha_2 \bigg[ 1 + \alpha_3 \; \exp \{\alpha_4 \; \alpha_5\} \bigg]^{-\frac{1}{\alpha_3}},
\end{equation}
and $\alpha_2 = -4.515523$, $\alpha_3 = 0.075651$, $\alpha_4 = -0.933822$, $\alpha_5 = -2.206251$, and $T_\text{c}$ is the temperature at the crest of the binodal surface, defined as:
\begin{equation}
    T_\text{c} = E \bigg( 1 + \frac{P}{D} \bigg),
\end{equation}
where $D = -35.0$~GPa and $E=4223.0$~K \citep{Stixrude2025}. The descending function, $g$ is itself a piecewise function:
\begin{equation}
    g(\tilde{y}) \left\{
  \begin{array}{lr} 
      g_1(\tilde{y}) & \tilde{y} > \tilde{y}_0 \\
      g_2(\tilde{y}) & \tilde{y} \leq \tilde{y}_0
      \end{array}
\right.  
\end{equation}
where $\tilde{y}_0=-5$ and $g_1$ shares its functional form with $f$:
\begin{equation}
    g_1(\tilde{y}) = \beta_1 + \beta_2 \bigg[ 1 + \beta_3 \; \exp \{- \beta_4 (\tilde{y} - \beta_5)\} \bigg]^{-\frac{1}{\beta_3}},
\end{equation}
where:
\begin{equation}
    \beta_1=\log_{10} T_\text{c} - \beta_2 \bigg[ 1 + \beta_3 \; \exp \{\beta_4 \; \beta_5\} \bigg]^{-\frac{1}{\beta_3}},
\end{equation}
and $\beta_2 = 0.544371$, $\beta_3 = 30.217687$, $\beta_4 = 2.504075$, $\beta_5 = -1.712032$. Finally, $g_2$ is log-linear:
\begin{equation}
    g_2(\tilde{y}) = 4\tilde{y} + \beta_0,
\end{equation}
where $\beta_0$ is given by:
\begin{equation}  \label{eq:analytic_Tb_last}
    \beta_0 = g_1(\tilde{y}_0) - 4 \tilde{y}_0.
\end{equation}

%%%%%%%%%%%%%%%%%%%%%%%%%%%%%%%%%%%%%%%%%%%%%%%%%%

% Don't change these lines
\bsp	% typesetting comment
\label{lastpage}
\end{document}